\documentstyle[12pt]{article}
\begin{document}
\pagestyle{empty}
\hfill{ANL-HEP-PR-94-59}
\vskip .5cm
\hfill{ITP-SB-94-41}
\vskip 1cm
\centerline{\Large \bf The Dilepton-Production Cross Section}
\centerline{\Large \bf in Principal Value Resummation}
\vskip .5cm
\begin{center}
\begin{tabular}{cc}
Lyndon Alvero$\dag$ & Harry Contopanagos\\
Institute for Theoretical Physics & High Energy Physics Division\\
SUNY Stony Brook & Argonne National Laboratory\\
Stony Brook, NY 11794-3840 & Argonne, IL 60439\\
\end{tabular}
\end{center}
\vskip 1cm
\centerline{August 1995}
\vskip .5cm
\centerline{\bf Abstract}

Using a recent calculation of the perturbative hard part for
dilepton production that sums large threshold corrections
to all orders in perturbative QCD, we compute the corresponding
cross sections. The hard part has been evaluated
using principal value resummation and contains all singular
momentum-dependent corrections. We also include a resummation
of large Sudakov terms, which are independent of parton momenta.
We give predictions for the dilepton-mass  distribution,
the rapidity distribution and the rapidity-integrated $K$-factor
at fixed-target energies and compare with various experimental
results in several kinematic regimes.
We find that principal value resummation produces cross sections
that are finite and well-behaved.  For both protons and  anti-protons
on fixed targets, the resummed cross sections are, in general, in excellent
agreement with the data.

\vskip 2.65in
\noindent
\hrulefill \\
$\dag${\small{\sl Present Address:  Physics Department, Pennsylvania
State University, 104 Davey Lab., University Park, PA 16802-6300}}

\newpage
\section{Introduction}
\pagestyle{plain}

It is well-known that certain hadron-hadron scattering cross sections,
evaluated in perturbative QCD (pQCD), suffer from the presence of
large corrections at all but very high energies \cite{ref:populace}.
These are associated with soft gluons that produce large threshold effects.
The primary example of such processes is the dilepton-production cross section,
in which the large corrections are associated with
the region where all the available partonic momentum is carried away by the
dilepton final state. We recently completed the calculation of  all these
large corrections \cite{ref:parti} to all orders in pQCD
based on the method of principal value resummation \cite{ref:paptwo},
whose main advantage is that it by-passes the infrared (IR) singularities
associated with the Landau pole of the QCD running coupling \cite{ref:six}.
The corresponding resummed hard part is thus independent of arbitrary
IR cutoffs, and can be used to produce unambiguous predictions
for the cross section.
These are, moreover, stable from a perturbative point of view, since
they include {\it all} such large corrections.

The principal value prescription exponentiates the corrections
into a leading and next-to-leading exponent, $E_L \ {\rm and} \ E_{NL}$,
respectively.  These exponents contain the effects of the one- ($E_L$) and
two-loop ($E_{NL}$) running coupling and asymptotically reproduce the resummed
perturbative series in the partonic phase space region where perturbation
theory is valid. Therefore, the corresponding resummed hard part is
well-defined mathematically throughout the whole kinematic region, with
all the correct properties associated with perturbation theory.
Furthermore, to completely calculate these exponents, which resum the
large threshold corrections to all orders in perturbation theory,
{\it only} one- and two-loop calculations of the corresponding hard
part are needed.  In \cite{ref:paptwo} and \cite{ref:parti},
we have performed an exhaustive study of the
properties of these exponents and the associated resummed hard
part in the context of the dilepton cross section. The main objective
of the present work is to use this hard part to compare predictions
for the physical cross section and related quantities with experiment.

In section 2 we will review the general formulation of the dilepton production
cross section, both with the resummed hard part and with
finite-order perturbative expressions up to two loops \cite{ref:vanNeervendis}.
Our formalism is valid for the differential cross section, with
respect to the dilepton invariant mass, as well as for its derivative
with respect to rapidity, $y$, at $y=0$.
In section 3 we review and discuss the resummation of
large Sudakov constants \cite{ref:three},  which are particularly
important as well at low invariant mass.  In section 4 we explicitly exhibit
the predictions resulting from our formalism, i.e.,
mass and mass-rapidity (or mass-$x_F$) distributions and $K$-factors,
and compare these with data from available fixed-target experiments.
We will find that principal value resummation produces corrections that are
finite and not very large compared to 2-loop corrections.  As one
goes to higher mass values, the gap between the resummed and finite-order
cross sections diminishes.  We will also see that the resummation is
sensitive to the choice of parton distributions, with global distributions
leading to larger cross sections than those obtained from older deep
inelastic scattering (DIS) fits.
Finally, in section 5 we summarize and discuss our findings.
Some numerical details are discussed in an appendix.
We postpone comparison with collider experiments to future work.

\section{The inclusive cross section for dilepton production}

In this section, we review the general formulation for the cross section for
dilepton production in hadronic collisions and provide
all the necessary formulas for our calculations.
This will streamline the discussion which we will give below and facilitate
comparisons with various experimental results.  For completeness, we present
the general formalism for the cross section, valid at either fixed-target
or collider energies, even though in this paper we will deal only with the
former, i.e., for invariant masses much smaller than the $W$ or $Z$ mass.
In the following, we will follow the notation of
\cite{ref:one}, \cite{ref:two}. In addition, we will also use the
notation that for any physical quantity of interest $R$, $R_L$
stands for this quantity calculated using principal value resummation with the
leading exponent $E_L$ only, $R$ for the same resummed quantity using the full
exponent $E=E_L+E_{NL}$, while in finite-order pQCD the same quantity
will be written as $R_i=\sum_{j=0}^i\alpha^jR^{(j)}$,
where $R^{(j)}$ is the $j$-th order perturbative correction and
$\alpha\equiv \alpha_s(Q^2)/\pi$.

In the reaction
\begin{equation}
h_1(p_1)+h_2(p_2)\to l{\bar l}(Q^\mu)+X,
\label{drellyan}
\end{equation}
the dilepton pair may be produced by
a  neutral virtual gauge boson $V$ ($V\in\{\gamma,\ Z\}$).
We will  use the standard notation $s=(p_1+p_2)^2,\  \tau=Q^2/s,$ where
$Q$ is the dilepton invariant mass.

\subsection{Mass distributions and $K$-factors}

We will first  deal with the single-differential inclusive cross section
\begin{equation}
{d\sigma^V\over dQ^2}=\sigma_B^V(Q^2)W^V(\tau,Q^2),
\label{singledif}
\end{equation}
where $\sigma_B^V$ is the point-like
Born cross section for $q\bar{q}\to V^*\to l\bar{l}$ and contains
all the electroweak parameters and physical dimensions.
In this paper, we are mostly concerned with large perturbative corrections
which come from the diagonal-flavor quark-antiquark subprocess.  However,
when we actually calculate the cross sections, we will also include the
contributions coming from the non-singular terms in the 1-loop hard part.
Furthermore, we will use parton distribution functions normalized to
deep-inelastic scattering experimental data (DIS scheme).

We may now write the hadronic factor as
\begin{eqnarray}
W^V(\tau,Q^2)&=&\sum_{a,b} C_{ab}^V\int_0^1{dx_1\over x_1}
{dx_2\over x_2}\omega_{ab}(\tau/x_1x_2,\alpha)
F_{a/h_1}(x_1,Q^2)F_{b/h_2}(x_2,Q^2) \nonumber\\
&=&\sum_{a,b} C_{ab}^V\int_\tau^1dz\ \omega_{ab}(z,\alpha){{\cal F}_{ab}
(\tau/z)\over z},
\label{wgamma}
\end{eqnarray}
where the sum extends over all active parton (quark, antiquark and gluon)
flavors,
\begin{eqnarray}
C_{f\bar f}^\gamma&=&C_{fg}^\gamma=C_{gf}^\gamma\equiv C_f^\gamma=e_f^2,
\nonumber \\
C_{f\bar f}^Z&=&C_{fg}^Z=C_{gf}^Z\equiv C_f^Z=
1+(1-4|e_f|{\rm sin}^2\theta_W)^2,
\label{substitute}
\end{eqnarray}
where $e_f$ is
the fractional charge of quark $f$ and $\theta_W$ is the electroweak mixing
angle.  Finally, the parton luminosity is defined as
\begin{equation}
{\cal F}_{ab}(\tau/z)=\int_0^1\delta(1-\tau/(zx_1x_2)){dx_1\over x_1}
{dx_2\over x_2}F_{a/h_1}(x_1,Q^2)F_{b/h_2}(x_2,Q^2).
\label{luminosity}
\end{equation}
For our purposes, it is convenient to integrate
eq.~(\ref{wgamma})\ by parts to obtain the equivalent expression
\begin{equation}
W^V(\tau, Q^2)=\sum_{a,b}C_{ab}^V\int_\tau^1dz\biggl[\int_z^1dz'
\omega_{ab}(z',\alpha)\biggr]{d\over dz}\biggl({{\cal F}_{ab}(\tau/z)
\over z}\biggr).
\label{wgammap}
\end{equation}

For the purely electromagnetic case, $V\equiv \gamma$, we have
\begin{equation}
\sigma_B^\gamma(Q^2)={4\pi\alpha_e^2\over 3NQ^2s},
\label{sigmagamma}
\end{equation}
where $N$ is the number of colors and $\alpha_e$ is the (electromagnetic) fine
structure constant.
Correspondingly, for $V\equiv Z$ we have
\begin{equation}
\sigma^Z_B(Q^2)={\pi\alpha_e \tau \over 4N{\rm sin}^2\theta_W
{\rm cos}^2\theta_WM_Z}\biggl[{\Gamma_{Z\to l{\bar l}}\over (Q^2-M_Z^2)^2+
M_Z^2\Gamma_Z^2}\biggr],
\label{sigmaz}
\end{equation}
where $M_Z= Z$ boson mass, $\Gamma_{Z\to l{\bar l}}$ is the partial decay
width of $Z$ to one dilepton species,
\begin{equation}
\Gamma_{Z\to l{\bar l}}={\alpha_eM_Z(1+(1-4\sin^2\theta_W)^2)\over
48\sin^2\theta_W\cos^2\theta_W},
\label{partialwidth}
\end{equation}
and  $\Gamma_Z$ is the total width of the $Z$ (summed over all leptonic and
hadronic decay channels).

The physical cross section $d\sigma/dQ^2$ will in general involve the sum of
$d\sigma^V/dQ^2$ over both vector bosons, as well as an interference term,
proportional to $\sigma_B^{\gamma Z}$, between the amplitudes mediated by a
photon and
a $Z$ \cite{ref:one}.  For dilepton production in the continuum, i.e., away
from resonances, which at fixed-target energies is\footnote{This is the
region where all fixed-target experiments are performed since for these
experiments, $s<M_Z^2$. For simplicity, we will exclude hadronic resonances,
like $J/\Psi$ and $\Upsilon$, which are subtracted from most experimental data
as well.} $Q^2\ll M_Z^2$, the quantities
$\sigma_B^Z,\ \sigma_B^{\gamma Z}$ are suppressed relative to $\sigma_B^\gamma$
by $(Q/M_Z)^4$ and $(Q/M_Z)^2$, respectively, and therefore we may write
\begin{equation}
{d\sigma\over dQ^2}\simeq {d\sigma^\gamma\over dQ^2}.
\label{sigmacont}
\end{equation}

The $K$-factor is a quantity that nicely measures the size
of the radiative corrections in the hard part by more-or-less cancelling
the effect of non-perturbative parton distributions.  It may be defined as
\begin{equation}
K\equiv {d\sigma/dQ^2\over d\sigma_0/dQ^2},
\label{kfactorgen}
\end{equation}
where the denominator is calculated in the same way as the numerator,
eqs.~(\ref{singledif}), (\ref{wgamma}), but using only the tree-level hard
part.
In the continuum, and in accordance with eq.~(\ref{sigmacont}), we may write
\begin{equation}
K\simeq K^\gamma={d\sigma^\gamma/dQ^2\over d\sigma_0^\gamma/dQ^2}={W^\gamma
(\tau,Q^2)\over W_0^\gamma(\tau,Q^2)}.
\label{kcontinuum}
\end{equation}
Since we will only be interested in fixed-target energies,
these are the formulas we will use throughout this paper.

We may now obtain resummed expressions for the above quantities
by using the resummed hard part $\omega_{ab}$ in the above formulas.
The expression for the resummed hard part is
\cite{ref:parti,ref:paptwo,ref:papzero}
$\omega_{ab}(z,\alpha)=\delta_{af}\delta_{b\bar f}\omega_{f\bar{f}}(z,\alpha)$
with
\begin{equation}
\omega_{f\bar{f}}(z,\alpha)=A(\alpha)I(z,\alpha),
\label{omegaf}
\end{equation}
where the function $A(\alpha)$ exponentiates the $\delta(1-z)$ terms
and will be analysed in detail in section 3,
while $I(z,\alpha)$ exponentiates the momentum-dependent plus-distributions
that produce large perturbative corrections at the edge of phase space
$z\to 1$. The form of $I(z,\alpha)$ depends on the finite-order structure
of the
plus-distributions to be resummed and, for best accuracy, should
reproduce that structure upon a Taylor expansion in $\alpha$.
In \cite{ref:parti}\ we generically studied $I(z,\alpha)$ using a
simplified expression
that
reproduces {\it most} of the finite-order large threshold corrections
up to two loops
and resums {\it the bulk} of these corrections to all orders.
That is sufficient for examining all the properties of the resummation.
Since in this work we will compare with experiment, however, we will use
a more accurate expression for the hard part, that reproduces {\it all}
of the finite-order large threshold corrections up to two loops and
resums them to all orders. Additional resummed structures in $I(z,\alpha)$,
that
are calculable in principal value resummation \cite{ref:paptwo,ref:papzero}
and would reproduce, upon expansion,
subdominant remaining threshold corrections starting at three-loop order,
are negligible.
For completeness, we reproduce the expression for
$I(z,\alpha)$ below:
\begin{equation}
I(z,\alpha)=\delta(1-z) -\left[{{\rm e}^{E}
\over (1-z)}\Gamma(1+P_1)Q[P_1,P_2]\right]_+
\label{hardpartf}
\end{equation}
with
\begin{equation}
Q[P_1,P_2]={\sin(\pi P_1)\over \pi}\biggl(1+P_2[\Psi^2(1+P_1)
+\Psi^{(1)}(1+P_1)-\pi^2]\biggr)+2P_2\cos(\pi P_1)\Psi(1+P_1)\ ,
\label{ques}
\end{equation}
where $E=E_L+E_{NL}$ contains all the large logarithmic corrections,
$\Gamma$ is the Euler Gamma function and $\Psi$, $\Psi^{(1)}$ are
the usual polygamma functions,
and
\begin{equation}
P_k(x,\alpha)\equiv {\partial^k E(x,\alpha)\over k! \partial x^k}\ ,
\label{pione}
\end{equation}
where it is understood that $E\equiv E(x,\alpha)$ is a function
that has a polynomial representation in $x\equiv \ln(1/(1-z))$ and $\alpha$
in the perturbative regime.  From eqs.~(\ref{wgammap}), (\ref{omegaf})
and (\ref{hardpartf}), we may
eliminate the $\delta$-function and plus-distributions by explicitly
integrating the former and using the property
$\int_z^1dz'[f(z')]_+=-\int_0^zdz'f(z')$ for the latter.
Hence, our final form for the resummed hadronic function is
\begin{equation}
W^V(\tau,Q^2)=\sum_f C_f^V A(\alpha)\int_\tau^1dz\left[1+
{\cal H}(z,\alpha)\right]{d\over dz}\left({{\cal F}_{f\bar{f}}(\tau/z)\over
z}\right),
\label{wgammapp}
\end{equation}
where the function ${\cal H}(z,\alpha)$ is given by
\begin{equation}
{\cal H}(z,\alpha)\equiv \int_0^{\ln({1\over 1-z})} dx
{\rm e}^{E\left(x,\alpha
\right)}
\Gamma\left(1+P_1\left(x,\alpha\right)\right)
Q[P_1(x,\alpha),P_2(x,\alpha)]\ .
\label{chard}
\end{equation}

It will  also be of interest to  compare the resummed expression for
the cross section or the $K$-factor with the corresponding quantities
obtained from finite-order calculations. The corresponding hadronic
factor is given by
\begin{equation}
W^V_i(\tau,Q^2)=\sum_{a,b}C_{ab}^V\int_\tau^1dz\biggl[\int_z^1dz'
\omega_{ab}^i(z',\alpha)\biggr]{d\over dz}\biggl({{\cal F}_{ab}
(\tau/z)\over z}\biggr),
\label{wgammapert}
\end{equation}
where the finite-order perturbative hard part is denoted by
\begin{equation}
\omega_{ab}^i(z,\alpha)=\sum_{j=0}^i\alpha^j\omega_{ab}^{(j)}(z).
\label{omegapert}
\end{equation}
The corresponding finite-order mass-distribution and $K$-factor are
given by
\begin{equation}
{d\sigma^V_i\over dQ^2}=\sigma^V_B(Q^2)W^V_i(\tau,Q^2)
\ \ {\rm and} \ \ K_i^V={W_i^V(\tau,Q^2)\over W_0^V(\tau,Q^2)}\ .
\label{sigmapert}
\end{equation}

The hard part $\omega_{f{\bar f}}^i$\ has been calculated up to second order
in both DIS and $\overline{\rm MS}$ schemes, in
\cite{ref:vanNeervendis}-\cite{ref:vanNeervendispp},
and \cite{ref:one}, respectively.
The full two-loop corrections
are quite lengthy but the leading corrections, as in the one-loop
case \cite{ref:eightla},  are the ``singular" ones in the $z\to 1$ limit,
i.e., the ones that are proportional to a delta function or a plus
distribution.  These come from virtual and soft-gluon corrections and are only
present in the $q\bar{q}$ channel of the initial partons.
We reproduce these corrections in the DIS scheme below, together with the
leading order hard part:
\begin{equation}
\omega_{ab}^{(0)} = \delta_{af}\delta_{b\bar f}\delta(1-z),
\label{omegzero}
\end{equation}
\begin{equation}
\omega_{f{\bar f}}^{(1)} =
{C_F \over 2}\biggl(2(1+z^2){\cal D}_1(z) + 3{\cal D}_0(z)
+ (1 + {4 \over 3}\pi^2)\delta(1-z) - 4z - 6\biggr),
\label{omegone}
\end{equation}
\begin{equation}
\omega_{fg}^{(1)} = \omega_{gf}^{(1)} =
{C_F \over 2}{3\over 8}\biggl(\biggl[z^2+(1-z)^2\biggr]\ln(1-z) + {9\over 2}
z^2 - 5z + {3\over 2}\biggr),
\label{omegoneqg}
\end{equation}
\begin{eqnarray}
\omega_{f{\bar f}}^{(2)}&\simeq&{1 \over 16}\biggl\{C_F^2
\biggl(32{\cal D}_3(z)+72{\cal D}_2(z)+[64\zeta(2)+52]
{\cal D}_1(z)+[112\zeta(3)+24\zeta(2)+15]{\cal D}_0(z)\biggr)\nonumber \\
& &+ C_A C_F\biggl(-44{\cal D}_2(z)+\biggl[{338 \over 9}-16\zeta(2)\biggr]
{\cal D}_1(z)+\biggl[57+{88 \over 3}\zeta(2)-24\zeta(3)\biggr]{\cal D}_0(z)
\biggr)\nonumber \\
& &+n_fC_F\biggl(8{\cal D}_2(z)-{44 \over 9}{\cal D}_1(z)-\biggl[
10+{16 \over 3}\zeta(2)\biggr]{\cal D}_0(z)\biggr)\nonumber \\
& &+\delta(1-z)\biggl(C_F^2\biggl[{548 \over 5}\zeta(2)^2+120\zeta(3)-3\zeta(2)
\biggr]\nonumber \\
& &+C_AC_F\biggl[{215 \over 9}+{2098 \over 9}\zeta(2)-{196 \over 3}\zeta(3)
-{154 \over 5}\zeta(2)^2\biggr]\nonumber \\
& &+ n_fC_F\biggl[{16 \over 3}\zeta(3)
-{340 \over 9}\zeta(2)-{38 \over 9}\biggr]\biggr)\biggr\},
\label{omegtwo}
\end{eqnarray}
where $\zeta(s)$ is the Riemann Zeta function and
\begin{equation}
{\cal D}_i(z)=\biggl[{\ln^i(1-z) \over 1-z}\biggr]_+.
\label{plustmp}
\end{equation}
The $\simeq$ in eq.~(\ref{omegtwo}) signifies that we
have dropped all terms that are non-singular in the $z\to 1$-limit. This,
of course, raises the question of how dominant the above singular terms are
relative to the non-singular ones, that occur both in the $q\bar{q}$ channel
and in the other initial-parton-state channels, or conversely, how important
are the contributions due to the non-singular terms.
A thorough numerical investigation of this issue has already been conducted
\cite{ref:one,ref:vanNeervendisp,ref:vanNeervendispp}.
The conclusions vary depending somewhat on the factorization scheme, and on the
range of energy and invariant mass.\footnote{In fact, for the
mass-distribution, the question can be fully answered up to two loops, given
the recent work of \cite{ref:vanNeervendispp}.  This is not so for the
differential mass distribution, described in sec. 2.2, since the full two-loop
results are not available at present.}  We may say, in general, that at
fixed-target energies, the 2-loop non-singular terms account for no more than
a 10\% {\it negative} contribution to the exact mass distribution over all
channels, with the approximation improving as $\tau$ increases.  For details
regarding variations of this result with the order, the factorization scheme,
etc., see \cite{ref:vanNeervendispp}.  Relative to the resummed
cross section then, the 2-loop non-singular terms would only lead to
a 3 to 7\%
reduction.  Given these general remarks, and because of the
complicated structure of the 2-loop non-singular contributions, we did not
include them in our cross section calculations but we should bear in mind
the reduction that they would induce in our resummed predictions.
As a last note, the terms proportional to $\delta(1-z)$ in eqs.~(\ref{omegone})
and (\ref{omegtwo}), will also be useful in determining the factor $A(\alpha)$
in eq.~(\ref{omegaf}), as will be shown in section 3.

\subsection{Differential mass distributions}

We can also use principal value resummation to compute the rapidity
distribution of dileptons in the central rapidity region.  This is possible
because the rapidity, or any other smooth function of the vector boson
momentum, when in the central region and expressed in terms
of the initial- and final-state momenta, is approximated by a function of the
initial parton momenta with corrections that are suppressed by vanishing
final-state phase space \cite{ref:laester}.
To be specific, it has been shown in \cite{ref:laester} that for any kinematic
variable ${\bar \eta}$ which is a smooth function of the vector boson momentum,
the resummed quantity $d^2\sigma^V / dQ^2d{\bar \eta}$ is approximated by
\begin{eqnarray}
{d^2\sigma^V\over dQ^2d{\bar \eta}}(\tau,Q^2,{\bar \eta})
&\simeq &\sigma^V_B{\displaystyle \sum_f} C_f^V
\int_0^1{dx_1\over x_1}{dx_2\over x_2}\omega_{f{\bar f}}(\tau/x_1x_2,
\alpha)F_{f/h_1}(x_1,Q^2)F_{{\bar f}/h_2}(x_2,Q^2)\nonumber \\
& & \times \ \delta({\bar \eta}-\eta(x_1p_1,x_2p_2))\equiv\widetilde{{
d^2\sigma^V\over dQ^2d{\bar \eta}}}(\tau,Q^2,{\bar\eta})\ ,
\label{generaldist}
\end{eqnarray}
where $\omega_{f{\bar f}}(z,\alpha)$ is given by eq.~(\ref{omegaf}) and
$\eta(x_1p_1,x_2p_2)$ denotes the corresponding parton-model expression.
By analogy to eqs.~(\ref{wgamma})
and (\ref{wgammap}), we can re-write eq.~(\ref{generaldist}) as
\begin{equation}
\widetilde{{d^2\sigma^V\over dQ^2d{\bar \eta}}}(\tau,Q^2,{\bar \eta})
\simeq \sigma^V_B{\displaystyle \sum_f}\
C_f^V \int_{\tau}^1 dz \biggl[\int_z^1 dz' \omega_{f{\bar f}}(z',\alpha)
\biggr]{d \over dz}\biggl({{\cal F}_{f{\bar f}}^{{\bar \eta}}(\tau/z)
\over z}\biggr),
\label{generaldistla}
\end{equation}
where
\begin{equation}
{\cal F}^{{\bar \eta}}_{f{\bar f}}(\tau/z) \equiv {\displaystyle \int_0^1}
{dx_1\over x_1}{dx_2\over x_2}\delta\left(1-{\tau/z\over x_1x_2}\right)
\delta({\bar \eta}-\eta(x_1p_1,x_2p_2))F_{f/h_1}(x_1)F_{{\bar f}/h_2}(x_2).
\label{lumineta}
\end{equation}

If we choose the kinematical variable ${\bar \eta}$ to be the rapidity $y$,
\begin{equation}
{\bar \eta}\equiv y={1\over 2}\ln\left({q_0+q_L\over q_0-q_L}\right),
\label{etar}
\end{equation}
we obtain for the partons
\begin{equation}
y(x_1p_1,x_2p_2)={1\over 2}\ln\left({x_1\over x_2}\right),
\label{ypartons}
\end{equation}
so that at $y=0$,
\begin{equation}
{d^2\sigma^V\over dQ^2dy}(\tau,Q^2,y=0)\simeq\sigma_B^VW^V_{y=0}(\tau,Q^2),
\label{rapid}
\end{equation}
where
\begin{equation}
W^V_{y=0}(\tau,Q^2)=\sum_f C_f^V A(\alpha)\int_\tau^1 dz[1+{\cal H}(z,\alpha)]
{d\over dz}\left({{\cal F}^{y=0}_{f{\bar f}}(\tau/z)\over z}\right),
\label{wyla}
\end{equation}
with
\begin{equation}
{\cal F}^{y=0}_{f{\bar f}}(\tau/z)=F_{f/h_1}\left(\sqrt{\tau/z}\right)
F_{{\bar f}/h_2}\left(\sqrt{\tau/z}\right).
\label{ffy}
\end{equation}

Similarly, choosing ${\bar \eta}$ to be the Feynman parameter $x_F$,
\begin{equation}
{\bar\eta}\equiv x_F\equiv 2q_L/\sqrt {s},
\label{exef}
\end{equation}
where $q_L$ is the longitudinal momentum of the virtual vector boson in the
center-of-mass frame of the hadrons, we have\footnote{In actual experiments,
$x_1$ typically corresponds to the momentum fraction of a parton in the beam
hadron while $x_2$ corresponds to that of a parton in the target hadron.}
\begin{equation}
x_F(x_1p_1,x_2p_2)=x_1-x_2\ .
\label{xfeynman}
\end{equation}
Therefore the resummed double-differential distribution at $x_F=0$ is
\begin{equation}
{d^2\sigma^V\over dQ^2dx_F}(\tau,Q^2,x_F=0)\simeq \sigma_B^V(Q^2)W_{x_F=0}^V
(\tau,Q^2),
\label{distone}
\end{equation}
where $W^V_{x_F=0}$ is given by eq.~(\ref{wyla}), with the substitution
\begin{equation}
{\cal F}^{y=0}(\tau/z)\rightarrow {\cal F}^{x_F=0}(\tau/z)={1\over
2\sqrt{ \tau/z}}F_{f/h_1}\left(\sqrt{\tau/z}\right)
F_{{\bar f}/h_2}\left(\sqrt{\tau/z}\right).
\label{ffx}
\end{equation}

Finite-order approximations can also be written down in a similar way:
\begin{equation}
{d^2\sigma^{V,i}\over dQ^2d{\bar \eta}}(\tau,Q^2,{\bar \eta})\simeq
\widetilde{{d^2\sigma^{V,i}\over dQ^2d{\bar\eta}}}(\tau,Q^2,{\bar\eta})\equiv
\sigma^V_B(Q^2)W^{V,i}_{{\bar \eta}}(\tau,Q^2),
\label{finQbeta}
\end{equation}
with
\begin{equation}
W^{V,i}_{{\bar \eta}}(\tau,Q^2)=\sum_{a,b} C_{ab}^V \int_\tau^1 dz\biggl[
\int_z^1 dz' \omega_{ab}^i(z',\alpha)\biggr]
{d\over dz}\left({{\cal F}^{{\bar \eta}}_{ab}(\tau/z)\over z}\right),
\label{finWv}
\end{equation}
where ${\bar \eta}=y \ {\rm or} \ x_F$ and $\omega_{ab}^i(z',\alpha)$
is given by eqs.~(\ref{omegzero})-(\ref{omegtwo}).

We illustrate the accuracy of the approximation (\ref{generaldist}) in
two ways.  First, we compare in fig.\ 1 the full ${\cal O}(\alpha)$
distribution
(i.e., all 1-loop non-singular terms in the hard part are included)
using eq.~(\ref{finQbeta}) (with $i=1$) and the exact distribution derived in
\cite{ref:eightla}, in the central region.
The kinematic parameters used are taken from the Fermilab E605 experiment
\cite{ref:esix} and are described in section 4.  The solid curve is the
exact one-loop calculation while the dotted curve is the corresponding
approximation. We see that the approximation is not perfect, with about a 10\%
difference between the two curves.  This difference comes mostly from
the quark-gluon channel.  Hence, in all our subsequent calculations for double
differential distributions, we used the exact results of \cite{ref:eightla}
for the quark-gluon channel alone.  The remaining error is then of order
1\%, which shows that the general approximation (eq.~(\ref{generaldist})) for
the channel containing the singular corrections is indeed very good.

In most experiments, what is measured is not the fully integrated mass
distribution $d\sigma/dQ^2$ but instead one that is {\it partially} integrated
over either the rapidity $y$ or $x_F$ in some kinematic range
${\cal R}(\tau)$.  In order to compare with actual experimental data,
we use the following approximation
\begin{equation}
\biggl({d\sigma^V \over dQ^2}\biggr)_{{\bar \eta}\in {\cal R}(\tau)}\equiv
\int_{{\bar \eta}\in {\cal R}(\tau)} d{\bar \eta}{d^2\sigma^V
\over dQ^2d\bar{\eta}} \simeq \int_{{\bar \eta}\in {\cal R}(\tau)}
d{\bar \eta}\widetilde {{d^2\sigma^V\over dQ^2d{\bar \eta}}}=
\sigma^V_B(Q^2)W^V_{{\bar \eta}\in {\cal R}(\tau)}(\tau,Q^2),
\label{appdsdQ}
\end{equation}
with $\widetilde{d^2\sigma^V}/dQ^2d\bar{\eta}$ given by eq.~(\ref{generaldist})
so that $W^V_{{\bar \eta}\in {\cal R}(\tau)}(\tau,Q^2)$ is given by
eq.~(\ref{wyla}) with the substitution
\begin{eqnarray}
{\cal F}_{f\bar{f}}^{y=0}(\tau/z)&\to& {\cal F}_{f\bar{f}}^{{\bar \eta}}
(\tau/z)\equiv\int_{{\bar \eta}\in{\cal R}(\tau)}d{\bar \eta}
\int_0^1{dx_1\over x_1}{dx_2\over x_2}\delta\left(1-{\tau/z\over x_1x_2}
\right)\nonumber \\
& &\times \delta({\bar \eta}-\eta(x_1p_1,x_2p_2))F_{f/h_1}(x_1)
F_{\bar{f}/h_2}(x_2).
\label{partint}
\end{eqnarray}
The second way, therefore, that we will test eq.~(\ref{generaldist}) is by
comparing eq.~(\ref{appdsdQ}) in finite-order pQCD with an exact calculation
at the end of this section.

The integrals over $x_1$, $x_2$ in eq.~(\ref{partint}) are trivial to perform.
Let us first consider the case of $\bar{\eta}\equiv x_F$
with ${\cal R}(\tau)$ the positive kinematic range $(0, 1-\tau)$,
as is the case for most experiments. We will denote the corresponding
function in eq.~(\ref{partint}) by ${\cal F}_{f\bar{f}}^{x_F>0}$.
Notice that upon performing the integrations over the parton fractions,
the constraints $x_1-x_2=x_F$ and $x_1x_2=\tau/z$ transform the
positive kinematic range of $x_F$ into $(0, 1-\tau/z)$, which
makes the function ${\cal F}_{f\bar{f}}^{x_F>0}$ depend on the scaling variable
$\tau/z$ only. We obtain the result
\begin{equation}
{\cal F}_{f\bar{f}}^{x_F>0}(\tau/z)=\int_0^{1-{\tau\over z}}{dx_F\over
\sqrt{x_F^2+4\tau/z}}
F_{f/h_1}\left({x_F+\sqrt{x_F^2+4\tau/z}
\over 2}\right)F_{\bar{f}/h_2}\left({-x_F+\sqrt{x_F^2+4\tau/z}\over 2}
\right).
\label{fxfpos}
\end{equation}
On the other hand, if $\bar{\eta}\equiv y$ and we want an analogous
expression which holds for $y>0$, the corresponding constraints are
$y={1\over 2}\ln(x_1/x_2)$ and $x_1x_2=\tau/z$ so that we get
\footnote{Note that $x_F>0$ implies $y>0$ and vice-versa, and in fact
using either $x_F$ or $y$ for representing the differential mass distribution
amounts to a change of integration variable, so that eqs.~(\ref{fxfpos})
and (\ref{fypos}) give identical results.   We found it convenient to use
eq.~(\ref{fypos}) for calculating the $y$ (or $x_F$) $>0$ cross sections.}

\begin{equation}
{\cal F}_{f\bar{f}}^{y>0}(\tau/z)=\int_0^{{1\over 2}\ln\left({z\over \tau}
\right)}dy F_{f/h_1}\left(\sqrt{{\tau\over z}}{\rm e}^y\right)
F_{\bar{f}/h_2}\left(\sqrt{{\tau\over z}}{\rm e}^{-y}\right)\ .
\label{fypos}
\end{equation}

Summarizing these results, we write
\begin{equation}
\biggl({d\sigma^V \over dQ^2}\biggr)_{{\bar \eta}>0}
\simeq \sigma^V_B(Q^2)W^V_{{\bar \eta}>0}(\tau,Q^2),
\label{appdsdQ2}
\end{equation}
where
\begin{equation}
W^V_{{\bar \eta}>0}(\tau,Q^2)=\sum_f C_f^V A(\alpha)\int_\tau^1 dz[1+{
\cal H}(z,\alpha)]
{d\over dz}\left({{\cal F}^{{\bar \eta}>0}_{f{\bar f}}(\tau/z)\over z}\right).
\label{wyla2}
\end{equation}
Finite-order approximations take a similar form:
\begin{equation}
\biggl({d\sigma^{V,i} \over dQ^2}\biggr)_{{\bar \eta}>0}\simeq
\sigma^V_B(Q^2)W^{V,i}_{{\bar \eta}>0}(\tau,Q^2),
\label{findsdQ}
\end{equation}
with $W^{V,i}_{{\bar \eta}>0}(\tau,Q^2)$ given by
eq.~(\ref{finWv}) but with ${\cal F}^{{\bar \eta}=0}_{ab}(\tau/z)$
replaced by ${\cal F}^{{\bar \eta}>0}_{ab}(\tau/z)$,
eq.~(\ref{partint}).

In fig.\ 2 we test the approximation shown in eq.~(\ref{appdsdQ}) by again
comparing the resulting partially integrated distribution, $\biggl({d\sigma^V
\over dQ^2}\biggr)_{y>0}$, with the exact one using the results of
\cite{ref:eightla}.  The calculations were done to one loop order, and
the comparison is made within the
context of the E772 experiment, with kinematics that we will describe in
sec. 4. We see that in this case, the approximation works very well.

In section 4 we will use these formulas in presenting our predictions
and comparisons with various fixed-target experiments.

\section{The function $A(\alpha)$}

We now turn to the coefficient function $A(\alpha)$ in eq.~(\ref{omegaf}),
which is closely connected to the Sudakov form factor.  In \cite{ref:six},
this function was given to ${\cal O}(\alpha)$ in the exponent as:
\begin{equation}
A_1(\alpha) = (1+2 C_F\alpha )\exp \biggl[{\alpha \over 2}C_F\biggr(
{4 \over 3}\pi^2-3\biggr) \biggr] .
\label{fcnA}
\end{equation}
In this expression, the piece ${\alpha \over 2 }C_F\pi^2$ of the exponent
is identified with
the absolute value squared of the ratio of the Sudakov form factor at
timelike and spacelike momentum transfer, $\Gamma(Q^2)/\Gamma(-Q^2)$.

To obtain an ${\cal O}(\alpha^2)$ form of $A(\alpha)$, we
use eq. (3.8) of \cite{ref:three}:
\begin{eqnarray}
\ln {\Gamma(Q^2) \over \Gamma(-Q^2)}&=&i{\pi \over 2}
\biggl[\alpha K^{(1)}+\alpha^2 K^{(2)} \biggr]
+{G^{(1)} \over 2 b_2}\ln \nu+{G^{(2)} \over 2 b_2}\alpha\left
(1-{1 \over \nu}\right)\nonumber \\
& &-{\gamma_K^{(1)} \over 4 b_2^2}{1 \over \alpha}
(\nu \ln \nu-\nu+1)
+{\gamma_K^{(2)} \over 4 b_2^2}(\ln \nu-\nu+1)
\label{lnratio}
\end{eqnarray}
where
\begin{eqnarray}
b_2&=& {11 \over 12}C_A-{2 \over 12}n_f,\nonumber \\
\nu&=& 1+i \pi b_2\alpha, \nonumber \\
\gamma_K^{(1)}&=& 2C_F,\nonumber \\
\gamma_K^{(2)}&=& C_AC_F\biggl[{67 \over 18}-\zeta(2)\biggr]-{5 \over 9}n_fC_F.
\label{gamcons}
\end{eqnarray}
The $\gamma_K^{(i)}$'s are anomalous dimension coefficients.

It is sufficient to consider only the real part of eq.~(\ref{lnratio}):
\begin{equation}
{\rm Re}\biggl[\ln {\Gamma(Q^2) \over \Gamma(-Q^2)}\biggr]
={G^{(1)} \over 2 b_2}\ln r+{\pi G^{(2)} \over 2}{\sin \theta \over r}
\alpha^2
-{\gamma_K^{(1)} \over 4 b_2^2}
{1 \over \alpha}(\ln r -\pi b_2 \theta\alpha)+
{\gamma_K^{(2)} \over 4 b_2^2}\ln r,
\label{relnratio}
\end{equation}
where
\begin{equation}
r=\biggl[1+(\pi b_2\alpha)^2\biggr]^{1 \over 2}\ ,\ \
\theta=\arctan(\pi b_2\alpha)\ .
\label{randtheta}
\end{equation}
The constant $G^{(1)}$ is obtained from eq. (2.7) of \cite{ref:three},
\begin{equation}
G^{(1)}={3 \over 2}C_F,
\label{geeup1}
\end{equation}
while the constant $G^{(2)}$ can be found from their eqs. (4.1), (4.2), (4.3),
(2.5) and (2.6)\footnote{There is a misprint in eq. (4.3) of \cite{ref:three}.
The right-hand side of the equation should read \\
$-4\epsilon\Gamma^{(2)}_2
\biggl[{\mu^2 \over -q^2}\biggr]^{2\epsilon}-2\epsilon\Gamma^{(2)}_1
\biggl[{\mu^2 \over -q^2}\biggr]^\epsilon-(K^{(1)}+G^{(1)})\Gamma^{(1)}$.}:
\begin{eqnarray}
G^{(2)}&=&C_F^2\biggl[{3 \over 16}-{7 \over 3}\zeta(3)+{23 \over 6}
\zeta(2)\biggr]
+C_AC_F\biggl[{2545 \over 432}+{11 \over 12}
\zeta(2)-{13 \over 4}\zeta(3)\biggr]\nonumber \\
& &+n_fC_F\biggl[-{209 \over 216}-{1 \over 6}\zeta(2)\biggr]\ .
\label{geeuptwo}
\end{eqnarray}
Expanding eq.~(\ref{relnratio})\ up to ${\cal O}(\alpha^2)$, one obtains
the exponentiated result
\begin{equation}
\biggl|{\Gamma(Q^2) \over \Gamma(-Q^2)}\biggr|^2=
\exp\left({\alpha \over 2}C_F\pi^2+{\alpha^2 \over 16 } \pi^2\biggl[\biggl(
{233 \over 9}-{2 \over 3}\pi^2\biggr)C_AC_F-{38 \over 9}n_fC_F\biggr]\right).
\label{ratexpan}
\end{equation}
The ${\cal O}(\alpha^2)$ exponentiated function $A(\alpha)$ can now
be written as
\begin{eqnarray}
A_2(\alpha)&=&\biggl(1+2C_F\alpha+b\alpha^2\biggr)\exp\Biggl(
{\alpha \over 2}C_F\biggl[{4 \over 3}\pi^2-3\biggr]\nonumber \\
& &+{\alpha^2 \over 16} \pi^2\biggl[\biggl({233 \over 9}-{2 \over 3} \pi^2
\biggr)C_AC_F-{38 \over 9}n_fC_F\biggr]\Biggr).
\label{fcnatwolp}
\end{eqnarray}
The constant $b$ is determined by expanding eq.~(\ref{fcnatwolp})\ in $\alpha$
and then comparing the result with the terms
proportional to $\delta(1-z)$ in the second-order result
$\omega_{f{\bar f}}^{(2)}$, eq.~(\ref{omegtwo}).
This yields
\begin{eqnarray}
b&=& C_F^2\biggl[{-23 \over 720}\pi^4-{35 \over 96}\pi^2
+{15 \over 2}\zeta(3)+{15 \over 8}\biggr]
+C_AC_F\biggl[{215 \over 144}+{175 \over 216}\pi^2
-{49 \over 12}\zeta(3)-{17 \over 1440}\pi^4\biggr]\nonumber \\
& &+n_fC_F\biggl[{1 \over 3}\zeta(3)-{19 \over 72}-{7 \over 54}\pi^2\biggr]\ .
\label{bcons}
\end{eqnarray}

There is yet another way of treating the function $A(\alpha)$.
The accuracy of the expansion of eq.~(\ref{relnratio})\ actually depends
on the size of $\pi b_2\alpha$.  Given the value of $b_2=2.0833 \ {\rm at}
\ n_f=4$, one can easily check that $\pi b_2\alpha$ is not really a very small
number.  Thus, instead of expanding in $\alpha$, one can use the exact
expression in eq.~(\ref{relnratio})\ to
obtain the following expression for \mbox {$A(\alpha)$:}\footnote{
In the following, it is this ``exact" expression that we will denote
by $A(\alpha)$ and use in the numerical computations.}
\begin{eqnarray}
A(\alpha)&=&\biggl(1+2C_F\alpha+b\alpha^2\biggr)\exp\Biggl(
{\alpha \over 2}C_F\biggl[{\pi^2 \over 3}-3\biggr]+{G^{(1)} \over b_2}\ln r
+\pi G^{(2)}{\sin \theta \over r}\alpha^2\nonumber \\
& &+{\gamma_K^{(1)} \over 2 b_2^2}{1 \over \alpha}(\pi b_2
\theta  \alpha-\ln r)+{\gamma_K^{(2)} \over 2 b_2^2} \ln r\Biggr)\ .
\label{fcnAexa}
\end{eqnarray}
Note that at high energies, eq.~(\ref{ratexpan})\ becomes a good enough
approximation to eq.~(\ref{relnratio})\ so that the constant $b$ in
eq.~(\ref{fcnAexa})\ is the same $b$ that appears in eq.~(\ref{fcnatwolp}),
given explicitly by eq.~(\ref{bcons}).

In fig.\ 3 we have plotted the different forms  $A_1(\alpha)$,
$A_2(\alpha)$ and $A(\alpha)$ for $n_f=4$ and $\Lambda=0.25{\rm GeV}$.
For consistency, we have used the two-loop form of $\alpha$ for $A_2(\alpha)$
and $A(\alpha)$ and a one-loop $\alpha$ for $A_1(\alpha)$.
One sees that the following relation holds: $A(\alpha)>A_2(\alpha)>A_1(\alpha)
$ \ where $A(\alpha)$ is given by eq.~(\ref{fcnAexa}), $A_2(\alpha)$ \ by
eq.~(\ref{fcnatwolp}) and $A_1(\alpha)$ \ by eq.~(\ref{fcnA}).
As $Q$ increases, all three functions become numerically close to 1, as
expected from asymptotic freedom. On the other hand,
at low values of $Q$ all three functions have large numerical values,
and diverge as $Q$ approaches the confinement scale.
In the mass-range of fixed-target experiments, all three
forms of $A$ are numerically very similar to one another.
In the following section we will give predictions for the resummed
dilepton cross section using only the exact, ${\cal O}(\alpha^2)$ function
$A(\alpha)$.

\section{Comparisons with experiment}

\subsection{Generalities}

In this section, we present our numerical calculations and compare them with
results from several fixed-target experiments.
To be consistent with the two-loop running coupling used in the calculation
of the resummed exponent \cite{ref:parti}, we use the following two-loop form
of $\alpha$ in terms of $\Lambda$:
\begin{equation}
\alpha(Q^2)\equiv {\alpha_s(Q^2)\over \pi}=
{1\over b_2\ln(Q^2/\Lambda^2)}-{b_3\over b_2^3}{\ln(\ln(Q^2/\Lambda^2))
\over \ln^2(Q^2/\Lambda^2)},
\label{twoloop}
\end{equation}
with
\begin{equation}
b_2=(11C_A-2n_f)/12, b_3=(34C_A^2-(10C_A+6C_F)n_f)/48.
\label{bees}
\end{equation}

For parton distributions, we used
the  recent
 CTEQ2D \cite{ref:cteq} and MRSD- \cite{ref:mrsd} parton sets that are
based on global fits.
In the hard part, we used the $\Lambda_{n_f}$ values provided with the
appropriate parton distribution.  The parameter $n_f$ is either 4 or 5
depending on the value of $Q$ at which the parton density is evaluated.

We note here that in all the discussions to follow, by resummed cross section,
we mean the resummed cross section obtained from eq.~(\ref{rapid}) or
(\ref{appdsdQ2}) plus the contributions of all the non-singular terms in the
1-loop hard part, $\omega_{ab}^{(1)}(z,\alpha)$.  Specifically, these are
$\alpha\omega_{fg}^{(1)}$ (eq.~(\ref{omegoneqg})) and
$\alpha{C_F\over 2}(-4z-6+2(1-z)\ln(1-z)-4\ln(1-z))$, derivable from
$\alpha\omega_{f\bar f}^{(1)}$ of eq.~(\ref{omegone}).

\subsection{The NA3 (1980) Experiment}

This experiment \cite{ref:nathree} involved the hadronic production
of dimuons from the interaction of a proton- $(\sqrt{s}=19.4$GeV) and
anti-proton- beam $(\sqrt{s}=16.8$GeV) with a platinum (Pt) target.
In fig.\ 4a we have plotted the data points together with the one-loop,
two-loop and resummed forms of the mass distribution $\displaystyle{M_{\mu
\mu}^3{d\sigma \over dM_{\mu\mu}}\biggl|_{x_F>0}}$
using the CTEQ2D ($\Lambda_4=0.235$GeV) distribution.
For both $p-$ and ${\bar p}-$ on Pt, all three curves exhibit acceptable
agreement with data.

We show in fig.\ 4b the resulting curves using the
MRSD$-$ ($\Lambda_4=0.23$GeV)
parton set.  For ${\bar p}$ on Pt, the curves are almost identical to the
corresponding ones in fig.\ 4a, while the $p$ on Pt curves are slightly
higher than their CTEQ2D counterparts.  These suggest that CTEQ2D and
MRSD$-$ have almost the same valence quarks but slightly different sea
quark parametrizations.

It should be noted here that the size of the resummed cross sections is
largely determined by the function $A(\alpha)$.  In fig.\ 5, by looking at the
dotted (Born) and dashed (resummed but with $A(\alpha)$ replaced by 1)
curves, we note that the contribution of the plus distributions steadily
increases with $\sqrt{\tau}$, from only a few percent in the range
$\sqrt{\tau}\leq 0.3$, up to about 26\% at $\sqrt{\tau}\simeq 0.5$.
Comparing the dashed curve with the solid one, we can
also determine that the plus distributions contribute from only a few percent
to about 11\%, in the same regions of $\sqrt{\tau}$, to the resummed cross
section.  One can also check that, in the mass range of this plot,
at least 80\% of the resummed cross section comes from $A(\alpha) \ \times $
Born cross section.

\subsection{The NA3 (1985) Experiment}

In this experiment \cite{ref:nathreep}, dimuon events, at $\sqrt{s}=27.4$ GeV,
produced in the collisions of protons with a platinum (Pt) target,
and within the range $x_F>0$ were analyzed.

In fig.\ 6 we present our resummed prediction for the mass distribution
$M^3_{\mu\mu}{\displaystyle{d\sigma\over dM_{\mu\mu}}}\bigg|_{x_F>0}$,
using the CTEQ2D ($\Lambda_4=0.235$GeV) parton set.
The resummed curve is no more than 23\% above the data, and the 1-loop
prediction fits the data best.

\subsection{The E537 Experiment}

This Fermilab experiment \cite{ref:efive} involved collisions of
an antiproton beam with a tungsten (W) target at $\sqrt{s}=15.4$ GeV.
Again, dimuon production events which fell in the positive $x_F$-range
were selected.

In fig.\ 7, we have plotted the data points and the mass distribution
$M^3_{\mu\mu}{\displaystyle{d\sigma\over dM_{\mu\mu}}}\bigg|_{x_F>0}$,
using the CTEQ2D parton distribution.
Here, the resummed curve has the best fit.  The 2-loop
curve also has a good fit but with a slight underestimate.

We have also plotted the differential mass distribution
$\displaystyle{M^3 {d^2\sigma\over dMdx_F}\bigg|_{x_F=0}}$
using the 1-, 2-loop and resummed hard parts in fig.\ 8.
The resummed curve has an excellent fit, while both finite order curves
underestimate the data.  Again, as in
the NA3(1980) ${\bar p}$ on Pt case, the error bars are so large
that all three curves are within them.

\subsection{The E605 Experiment}

In this Fermilab experiment \cite{ref:esix}, a proton beam was incident on
a copper (Cu) target at $\sqrt{s}=38.8$ GeV. We show in figs.\ 9a and 9b
the data points with the 1-loop, 2-loop and resummed predictions for
$\displaystyle{s {d^2\sigma \over d{\sqrt \tau}dy}\biggl|_{y=0}}$ using the
CTEQ2D and MRSD- distributions, respectively.

E605 data together with a 1-loop hard part as a theoretical input, were used
for the CTEQ global fit, so, as expected, the 1-loop curve in fig.\ 9 gives
an excellent fit to the data over the whole range of $\tau$.  The resummed
cross section is about 18 to 14\% bigger than the data, but within
experimental errors,
the difference decreasing on average as $\tau$ increases.

In fig.\ 9b, the 1-loop curve also has the best fit to the data but with
some overestimate.  The 2-loop and resummed curves overestimate the data,
and are about 10\% bigger than their CTEQ2D counterparts in fig.\ 9a.

\subsection{The E772 Experiment}

This recent Fermilab experiment \cite{ref:eseven} focused on interactions of a
proton beam with various targets, to obtain, among other things, the
nuclear dependence of the resulting spectrum. Here we will look at
$p$-$^2$H cross sections at $\sqrt{s}=38.8$ GeV and $x_F>0$.

In fig.\ 10, we plot the mass distribution
$\displaystyle{M_{\mu\mu}^3{d\sigma\over dM_{\mu\mu}}}\biggl|_{x_F>0}$
using the CTEQ2D parton distributions.
In the low mass region $(.1< \sqrt{\tau} < .24)$, the resummed curve has
the best fit to the data.
In the high mass region $(.1< \sqrt{\tau} < .24)$, the 1-loop curve has the
best fit.  However, the resummed and 2-loop curves are no more than 16\%
bigger than the 1-loop curve, and within experimental errors.

We show in fig.\ 11 the distribution
$\displaystyle{M_{\mu\mu}^3{d^2\sigma\over dM_{\mu\mu}dx_F}}\biggl|_{x_F=0}$
using CTEQ2D parton distributions.  In the high mass region, all three curves
pass through almost all of the error bars.
The resummed and 2-loop curves, which in this case are almost identical, have
the best fit.  In the low mass region, the 2-loop
cross section gives a good fit, with some slight underestimate.  The resummed
curve overshoots the data by 18\% or less.

\subsection{$K$-factors}

We have also calculated a set of K-factors
for the kinematics of
various experiments using eqs.~(\ref{kcontinuum}), (\ref{wgammapp}),
(\ref{sigmapert}) and (\ref{wgammapert}).
Note that the denominator is simply the Leading
Logarithm Approximation (LLA) cross section.
For figs. 12 and 14, the
$K$ factors were calculated over the full rapidity range, while for fig.\ 13,
the $K$'s were only calculated over positive rapidity.
These quantities express the
size of the resummed radiative corrections relative to lowest order.
They turn out to be relatively insensitive to the parton sets used, and allow
us to compare the importance of corrections for the various experiments.

In fig.\ 12, we plot $K_1$, $K_2$ and the resummed $K$ factor, using
parameters from the E537 experiment described above and the CTEQ2D parton
distributions.  $K_i$ is the $i$-th loop $K$-factor obtained using
eq.~(\ref{sigmapert}).  To better illustrate the different contributions to
$K$, we show two other curves in the same plot:  the dot-dashed and dotted
curves are resummed curves but with $A(\alpha)$ (eq.~(\ref{fcnAexa})) replaced
by $A_1(\alpha)$ (eq.~(\ref{fcnA})) and $1$, respectively.

We see that, in the mass range covered by the plot, most of the resummed $K$
comes from the $A(\alpha)$ term, as the plus distributions in this case
account from only a few percent to about 26\% of the total $K$.
This agrees with an earlier
assessment given in the cross section discussions.  Comparing the solid
and dot-dashed curves, one also observes that the next-to-leading terms in
the exponent of $A(\alpha)$ yield from 3 to 7\% contribution to the resummed
cross sections.

In fig.\ 13, we present our theoretical resummed $K$-factor, along with
$K_1$ and $K_2$ for the NA3 (1985) experiment, using two different parton
sets: CTEQ2D and MRSD$-$.   We notice that the resulting curves
are in very good agreement with one another.

In fig.\ 14, we present the $K$-factors, using the CTEQ2D parton distributions
for the E772 experiment.  What is important to note here is that as one goes
to large mass values, the difference between the resummed and the 2-loop
$K$-factor decreases to about a few percent.  Moreover,
whereas $K_1$ and $K_2$ are noticeably increasing in this region and beyond,
$K$ is relatively flat or very smoothly decreasing.  This highlights
the ``well-behaved'' nature of principal value resummation, for it is precisely
in this region where the logarithms in perturbative calculations are large.

\subsection{Theoretical Uncertainty}

In this section, we quantify the theoretical uncertainty in our resummed
calculations, as well as the goodness-of-fit of both resummed and finite-order
cross sections.  For purposes of discussion, we focus on data from the E605
and E537 experiments.

We compare in fig.\ 15 the resummed cross sections calculated using CTEQ2D
and MRSD$-$ parton distributions with data from E605.  We see that
the 2 curves are within 10\% of each other.
There is also an estimated uncertainty of about 7\% due to contributions of
non-singular terms that were not included in the resummed hard part (see
discussion in section 5).  Thus, we consider a good estimate of the theoretical
uncertainty in the resummed predictions to be 17\%.
%Note that in this estimate, we did
%not include our other resummed results using the DIS-based parton sets that
%were fitted to older data.
%This issue of combining the resummed hard part with
%DIS-based parton sets is addressed more fully in section 5.

In table 1, we present the $\chi^2$ per degree of freedom of the
calculated cross sections relative to data from E605.
As expected, the 1-loop cross sections calculated with global parton sets have
the best $\chi^2$ values.  However, one also notes that the $\chi^2$'s of the
resummed cross sections are not that large compared with the best values.
\footnote{In \cite{ref:LA}, we discuss the possibility that this difference is
related to nonperturbative effects.}

\begin{center}
\begin{tabular}{||cccc||}
\multicolumn{4}{c} {TABLE 1} \\ \hline
\multicolumn{4}{|c|}{$\chi^2$ per degree of freedom (E605)} \\ \hline\hline
Parton set & Resummed & 1-loop & 2-loop \\ \cline{1-4}
CTEQ2D & 5.0 & 1.3 & 2.1 \\ \hline
MRSD- & 11.6 & 2.6 & 5.9 \\ \hline
\end{tabular}
\end{center}

Similarly, we present in table 2 the analogous $\chi^2$ values for the
experiment E537, using the data points shown in our fig.\ 8.  In this case,
the resummed cross sections have the best $\chi^2$ values, while the
finite-order cross sections have $\chi^2$'s not more than 1.
This supports our earlier claim that for antiprotons on fixed targets the
resummed cross section has the best fit.

\begin{center}
\begin{tabular}{||cccc||}
\multicolumn{4}{c} {TABLE 2} \\ \hline
\multicolumn{4}{|c|}{$\chi^2$ per degree of freedom (E537)} \\ \hline\hline
Parton set & Resummed & 1-loop & 2-loop \\ \cline{1-4}
CTEQ2D & 0.05 & 0.38 & 0.84 \\ \hline
MRSD- & 0.14 & 0.46 & 0.93 \\ \hline
\end{tabular}
\end{center}

\section{Discussion and Conclusions}

In this paper, we have used a previously calculated hard part \cite{ref:parti}
for dilepton production, which resums large threshold corrections
to all orders in pQCD via the method of principal value resummation
\cite{ref:paptwo}, to calculate resummed DY distributions.
We have also derived the coefficient function $A(\alpha)$, which
involves the resummation of large, $z$-independent Sudakov terms,
to next-to-leading order.  We then made comparisons of several distributions
with data from fixed-target experiments.  In as much as the available
data cover only the positive rapidity region, we used the approximation of
\cite{ref:laester} in our
calculations. All calculations have been made in the DIS scheme.
To be entirely consistent, one
should use parton distributions resulting purely from  fits
to Deep Inelastic Scattering data. Existing DIS fits
(like DFLM \cite{ref:dflm} and EHLQ \cite{ref:ehlq}),
however,
had been constructed from relatively old data.
On the other hand, the global fits (like CTEQ2D and MRSD$-$) that we have
used, typically
use only 1-loop hard parts for input in hadron-hadron scattering, for example
when Drell-Yan data is included, but include the most recent DIS data.
If higher order corrections are important,
their effects will be included in the globally fitted distributions,
and using them in a resummed calculation would involve some double-counting
and thus lead to overestimates.
%Given the choices, we still think
%using parton sets with recent DIS data is more appropriate, since
%this data consitutes the ``bulk" of the fits anyway.

We now summarize our results in the light of these issues.
For all fixed-target experiments we have examined, the resummed
cross sections exhibit good to excellent agreement with the data.
For antiprotons on fixed targets, we find in general that the resummed
predictions agree better than the finite-order ones.
For proton beams the finite-order predictions agree better in most cases,
but in those cases the resummed ones are above the data by no more than
$10-23\%$
and even then, all predictions are within experimental errors.
 From the above, we can conclude that principal value resummation has
successfully ``tamed'' the divergences in perturbation theory associated with
the resummation of plus-distributions.  We find
that even if large corrections exist at the 1- and 2-loop order,
principal value resummation yields
cross sections that are finite and well-behaved. In addition, the
quantitative structure of the large threshold corrections in principal
value resummation correctly accounts for the observed data.

It is important here to summarize what these results represent concerning
our knowledge of perturbative QCD.
Our resummed cross sections were calculated including all the radiation
phase space down to threshold. This includes the perturbative regime,
which provides the bulk of the effect, and the non-perturbative regime
which we treat via the principal value resummation model.

The fact that, using global parton sets, the resummed cross section slightly
overshoots most proton beam data, needs to be disentangled:

It would be useful if
a parton set fitted only to the most recent DIS data becomes available,
so that one could repeat our resummed calculations with this set.
If, however, the antiquark fits are so ``robust" that the resulting cross
sections do not change significantly then the following issues, which could
possibly explain the overestimate, need to be confronted.
First, the cross section could be sensitive to higher twist.  The physical
higher-twist would be slightly different than in principal value resummation,
and should be parametrized more generally.
This issue is studied in more detail in \cite{ref:LA}.

Secondly, the factor $A(\alpha)$ accounts for a large part of the
resummed cross section. It is important to note that only the
exponentiation of Sudakov terms in $A(\alpha)$ is fully understood
at this moment. Since the exponent of eq.~(\ref{fcnAexa}) contains some finite
non-Sudakov terms as well, following \cite{ref:gs87},
a better understanding of the resummation of these
terms in this function is also needed to eliminate some of the
theoretical uncertainty.  It might be that inclusion of such terms
might make $A(\alpha)$ smaller, so that the resummed cross section is also
reduced.

Another issue is the size of the  non-singular terms in the hard
part at higher orders.
For the case of the 2-loop order, the size
of the contributions of these non-singular terms has been studied.  From
\cite{ref:vanNeervendispp}, we note that in the kinematic region
we have considered, the non-singular terms in the 2-loop hard part
yield from 5 to 10\% negative contributions to the cross section.  In the
resummed case, those terms will lead up to a $7\%$ negative
contribution.\footnote{Note that we did not include the 2-loop non-singular
contributions in our central value predictions, but we did so in our
uncertainty estimate.}
If such ``finite'' terms
continue to be sizeable at still higher orders, however, they might further
reduce the
effects of resummation, which might explain why 1- or 2-loop cross sections
that use global parton sets can describe the data quite well.

All of these issues are correlated in the differences
of our predictions using different parton sets, and the 2-loop reduction
of the non-singular terms. Our uncertainty estimate of $17\%$
is in fact a conservative measure of these factors and clearing-up these
issues corresponds to narrowing down this theoretical uncertainty.

In future work we hope to extend these calculations in the context of collider
energies, both for the Drell-Yan continuum and for vector boson production.
Among the most interesting other reactions sharing many common features
with dilepton production, and where principal value resummation has
been advantageously applied, is top-quark production \cite{ref:top}.

%\newpage

\centerline{ACKNOWLEDGMENTS}

We  thank George Sterman for many valuable suggestions, and Ed
Berger, Eric Laenen, Jeff Owens, Jack Smith  and Wu-Ki Tung for many helpful
discussions.   This work was supported in part by the U.S. Department of
Energy, Division of High Energy Physics, contract W-31-109-ENG-38, and by the
National Science Foundation,  grant PHY 9309888.

%\newpage

\appendix{\bf APPENDIX A Some Notes on the Numerical Calculations}

The explicit expression used to calculate the resummed cross section
in this paper is
\begin{equation}
{d\sigma \over dQ^2}=\sigma_B\sum_f e_f^2\int_{
\tau}^1 dz A(\alpha) [1+{\cal H}(z,\alpha)]{d\over dz}
\left({{\cal F}_{f{\bar f}}(\tau/z)\over z}\right)\ ,
\label{defcs}
\end{equation}
where $\sigma_B$ is given by eq.~(\ref{sigmagamma}), $A(\alpha)$ by
eq.~(\ref{fcnAexa}), ${\cal H}(z,\alpha)$
by eq.~(\ref{chard}) and the parton luminosity by eq.~(\ref{luminosity}).
The factor involving the luminosity can be expressed as an integral over the
rapidity $y$ \cite{ref:six}:
\begin{equation}
{d\over dz}\biggl({{\cal F}_{f\bar{f}}(\tau/z)\over z}\biggr)={\displaystyle
-{1 \over 2z^2}\int_{y^-}^{y^+} dy \biggl(2+x_1{d \over
dx_1}+x_2{d \over dx_2}\biggr)}F_{f/h_1}(x_1,Q^2)F_{\bar{f}/h_2}(x_2,Q^2),
\label{derlumy}
\end{equation}
where $x_1={\sqrt {\tau \over z}}{\rm e}^y,\ x_2={\sqrt {\tau \over z}}{\rm
e}^{-y} \ {\rm and}\ y^{\pm}=\pm {1\over 2} \log ({z \over \tau})$.

Note that eq.~(\ref{defcs}) may now be re-written as
\begin{eqnarray}
{d\sigma \over dQ^2}&=& A(\alpha){d\sigma_0 \over dQ^2}
+ A(\alpha)\sigma_B\sum_f e_f^2\int_{\tau}^1 dz{\cal H}(z,\alpha)
\biggl(-{1 \over 2z^2}\biggr)\int_{y^-}^{y^+}dy \biggl(2+x_1{d
\over dx_1}+x_2{d \over dx_2}\biggr)\nonumber \\
& &\times F_{f/h_1}(x_1,Q)F_{{\bar f}/h_2}(x_2,Q),
\label{defcs2}
\end{eqnarray}
with $d\sigma_0 \over dQ^2$ as in eq.~(\ref{kfactorgen}).
With the integral definition of ${\cal H}(z,\alpha)$, eq.~(\ref{chard}),
one has to perform at least four integrations to compute $d\sigma /dQ^2$
for one value of $Q$.\footnote{Aside from the $z,z' \ {\rm and} \
y$ integrals, further integrations are required to calculate $E$ and $P_1$.}
Straightforward implementation of eq.~(\ref{defcs})\ is thus very
time-consuming from a numerical point of view.

In order to shorten significantly the processing time requirements, an
intermediate step was performed: the quantity
${\cal H}(z,\alpha)$ was first interpolated over a range of $z$ and $\alpha$
values.  The interpolation is essentially a two parameter ($z,\alpha$) fit to
a set of ${\cal H}(z,\alpha)$ values calculated over the box contour for
particular values of $z$ and $\alpha$.  But because $E$ (and thus $P_1$)
\footnote{In calculating the set of ${\cal H}(z,\alpha)$ values, $E$ was
calculated over a box contour \cite{ref:paptwo}
while $P_1$ was calculated numerically:
$P_1(z)=(1-z){E(z+h)-E(z-h) \over 2h}$ for some small number $h$.} depends on
parameters which in turn depend on $n_f$, there is an implicit dependence of
${\cal H}(z,\alpha)$ on $n_f$.  Furthermore, note that one can have a leading
or next-to-leading ${\cal H}(z,\alpha)$ depending on whether $E=E_L\ {\rm or}
\ E=E_L+E_{NL}$ is used.  For all the fixed target cross section calculations,
$n_f=4\ {\rm and}\ E=E_L+E_{NL}$ were used.

Armed with the interpolating function, the calculation of eq.~(\ref{defcs2})
becomes a little simpler (for now, only a $2$-fold integral is involved)
and more importantly, a lot faster from a numerical point of view.

The interpolating function, which we will now denote by
${\cal H}_I(z,\alpha)$, \footnote{The interpolations were obtained using the
B spline routines in the Fortran 77 IMSL package.} is valid for \\
$.01<z<1-10^{-6} \ {\rm and} \ 4.756\times 10^{-2}<\alpha<7.197\times 10^{-2}$.
The choice of the boundaries for $z$ is essentially machine-dependent:
outside this range in $z$, the program ran into round-off errors.  Hence,
when ${\cal H}_I(z,\alpha)$ is used in eq.~(\ref{defcs2}), the upper
$z$-limit is taken to be $1-10^{-6}$.  As mentioned in our previous paper
\cite{ref:parti}, for fixed $\alpha$, the exact ${\cal H}(z,\alpha)$ is
bounded, reaching a certain limiting value before $z\rightarrow 1-10^{-6}$.
The contribution from the region $z>1-10^{-6}$ is thus expected to be very
small and may be neglected.

For the finite order calculations, instead of
eq.~(\ref{defcs}), (\ref{derlumy}), we use
\begin{eqnarray}
{d\sigma \over dQ^2}^i=&\sigma_B{\displaystyle \sum_{j=0}^i\sum_{a,b}
C_{ab}^V\int_{
\tau}^1 dz \biggl[\int_z^1 dz' \alpha^j \omega^{(j)}_{ab}
(z',\alpha)\biggr]\biggl(-{1 \over 2z^2}\biggr)\int_{y^-}^{y^+}
dy \biggl(2+x_1{d \over dx_1}+x_2{d \over dx_2}\biggr)}\nonumber \\
&\times{\displaystyle F_{a/ h_1}(x_1,Q)
F_{b/h_2}(x_2,Q)},
\label{defcsp}
\end{eqnarray}
with $C_{ab}^V$ given in section 2.1 and
$\omega^i_{ab}(z',\alpha)$ given by eqs.~(\ref{omegzero})-(\ref{omegtwo}).
No interpolations were necessary for the finite-order hard-parts.

For the exact calculation of ${d\sigma^{(1)}_{ab} \over dQ^2dy}|_{y=0}$,
we used the results given in \cite{ref:eightla}.  The explicit
expression for the $q\bar q$ channel is
\begin{equation}
{d\sigma^{(1)}_{q{\bar q}}\over dQ^2dy}\biggl|_{y=0}={16\pi \over 27}
{\alpha_e^2\alpha\over Q^2s}\sum_fe_f^2\sum_{j=1}^5 T^j_{f\bar f},
\label{fx1lpQy}
\end{equation}
with
\begin{eqnarray}
T^1_{f\bar f}&=&{\textstyle {1\over 2}\biggl[1+{5\over 3}\pi^2-{3\over 2}
\ln {x_1x_2\over (1-x_1)(1-x_2)}+2\ln {x_1\over (1-x_1)}\ln {x_2\over (1-x_2)}
\biggr]P_{f\bar f}(x_1,x_2,Q)},\nonumber \\
T^2_{f\bar f}&=&{\textstyle {1\over 2}\int_{x_1}^1dt_1\biggl({t_1^2+x_1^2\over
t_1^2(t_1-x_1)_+}\ln {2x_1(1-x_2)\over x_2(t_1+x_1)}+{3\over 2}{1\over
(t_1-x_1)_+}-{2\over t_1}-{3x_1\over t_1^2}\biggr)P_{f\bar f}(t_1,x_2,Q)},
\nonumber \\
T^3_{f\bar f}&=&{\textstyle {1\over 2}\int_{x_2}^1dt_2\biggl({t_2^2+x_2^2\over
t_2^2(t_2-x_2)_+}\ln {2x_2(1-x_1)\over x_1(t_2+x_2)}+{3\over 2}{1\over
(t_2-x_2)_+}-{2\over t_2}-{3x_2\over t_2^2}\biggr)P_{f\bar f}(x_1,t_2,Q)},
\nonumber \\
T^4_{f\bar f}&=&{\textstyle \int_{x_1}^1dt_1\int_{x_2}^1dt_2{G^A(t_1,t_2)\over
[(t_1-x_1)(t_2-x_2)]_+}P_{f\bar f}(t_1,t_2,Q)},\nonumber \\
T^5_{f\bar f}&=&{\textstyle \int_{x_1}^1dt_1\int_{x_2}^1dt_2{-2\tau (\tau+t_1
t_2)\over t_1t_2(t_1x_2+t_2x_1)^2}P_{f\bar f}(t_1,t_2,Q)},
\label{fxts}
\end{eqnarray}
where ${\textstyle x_1=x_2={\sqrt \tau}}$,
${\textstyle P_{f\bar f}(t_1,t_2,Q)=F_{{f/h_1}}(t_1,Q^2)F_{{{\bar f}/h_2}}
(t_2,Q^2)}$
and
\begin{equation}
G^A(t_1,t_2)={(\tau+t_1t_2)(\tau^2+(t_1t_2)^2)\over
(t_1t_2)^2(t_1+x_1)(t_2+x_2)}.
\label{t4ga}
\end{equation}
The sum in eq.~(\ref{fx1lpQy}) extends over all active quark and antiquark
flavors.  Eqs.~(\ref{fx1lpQy}) and (\ref{fxts}) were obtained using eqs.~
(2.1), (2.3), (2.6)-(2.8) of \cite{ref:eightla}.  The $+$ distributions are
evaluated using the prescription given by their eqs. (2.9) and (2.10).

Similarly, for the quark-gluon channel, we used eqs. (2.13)-(2.17) of
\cite{ref:eightla} to obtain
\begin{equation}
{d\sigma^{(1)}_{qg}\over dQ^2dy}\biggl|_{y=0}={2\pi \over 9}
{\alpha_e^2\alpha\over Q^2s}\sum_fe_f^2\sum_{j=1}^6 U^j_{gf},
\label{fx1lpQyqg}
\end{equation}
with
\begin{eqnarray}
U^1_{gf}&=&{\textstyle \int_{x_1}^1dt_1\biggl({x_1^2+(t_1-x_1)^2\over 2t_1^3}
\ln{2x_1(1-x_2)\over x_2(t_1+x_1)}+{1\over 2t_1}-{3x_1(t_1-x_1)\over t_1^3}
\biggr)P^C_{gf}(t_1,x_2,Q)},\nonumber \\
U^2_{gf}&=&{\textstyle \int_{x_1}^1dt_1\int_{x_2}^1dt_2{G^C(t_1,t_2)\over
(t_2-x_2)_+}P^C_{gf}(t_1,t_2,Q)},\nonumber \\
U^3_{gf}&=&{\textstyle \int_{x_1}^1dt_1\int_{x_2}^1dt_2H^C(t_1,t_2)
P^C_{gf}(t_1,t_2,Q)},\nonumber \\
U^4_{gf}&=&{\textstyle \int_{x_2}^1dt_2\biggl({x_2^2+(t_2-x_2)^2\over 2t_2^3}
\ln{2x_2(1-x_1)\over x_1(t_2+x_2)}+{1\over 2t_2}-{3x_2(t_2-x_2)\over t_2^3}
\biggr)P^{C'}_{gf}(t_2,x_1,Q)},\nonumber \\
U^5_{gf}&=&{\textstyle \int_{x_2}^1dt_2\int_{x_1}^1dt_1{G^C(t_2,t_1)\over
(t_1-x_1)_+}P^{C'}_{gf}(t_2,t_1,Q)},\nonumber \\
U^6_{gf}&=&{\textstyle \int_{x_2}^1dt_2\int_{x_1}^1dt_1H^C(t_2,t_1)
P^{C'}_{gf}(t_2,t_1,Q)},
\label{fxus}
\end{eqnarray}
where
\begin{eqnarray}
G^C(t_1,t_2)&=&{\textstyle {x_2(\tau+t_1t_2)(\tau^2+(\tau-t_1t_2)^2)\over
t_1^3t_2^2(t_1x_2+t_2x_1)(t_2+x_2)}},\nonumber \\
H^C(t_1,t_2)&=&{\textstyle {\tau(\tau+t_1t_2)(t_1t_2^2x_1+\tau(t_1x_2+2t_2x_1))
\over(t_1t_2)^2(t_1x_2+t_2x_1)^3}},\nonumber \\
P^C_{gf}(t_1,t_2,Q)&=&{\textstyle F_{g/h_1}(t_1,Q^2)F_{f/h_2}(t_2,Q^2)},
\nonumber \\
P^{C'}_{gf}(t_2,t_1,Q)&=&{\textstyle F_{g/h_2}(t_2,Q^2)F_{f/h_1}(t_1,Q^2)}.
\label{fxusgh}
\end{eqnarray}

For the exact calculation of ${d\sigma^{(1)}_{q{\bar q}}\over dQ^2}
\biggl|_{y>0}$, we used
\begin{equation}
{d\sigma^{(1)}_{q{\bar q}}\over dQ^2}\biggl|_{y>0}={16\pi \over 27}
{\alpha_e^2\alpha\over Q^2s}\sum_fe_f^2\sum_{j=1}^5 S^j_{f\bar f},
\label{fx1lpQ}
\end{equation}
with
\begin{eqnarray}
S^1_{f\bar f}&=&{\textstyle{1\over 2}\int_{{\sqrt \tau}}^1{dx_1\over x_1}
\biggl(1+{5\over 3}\pi^2-{3\over 2}\ln {\tau\over (1-x_1)(1-\tau/x_1)}+2
\ln {x_1\over (1-x_1)}\ln {\tau\over (x_1-\tau)}\biggr)P_{f\bar f}(x_1,
\tau/x_1,Q)},\nonumber \\
S^2_{f\bar f}&=&{\textstyle {1\over 2}\int_{{\sqrt \tau}}^1{dx_1\over x_1}
\int_{x_1}^1dt_1\biggl({t_1^2+x_1^2\over t_1^2(t_1-x_1)_+}\ln {2x_1(x_1-\tau)
\over \tau(t_1+x_1)}+{3\over 2}{1\over (t_1-x_1)_+}-{2\over t_1}-
{3x_1\over t_1^2}\biggr)P_{f\bar f}(t_1,\tau/x_1,Q)},\nonumber \\
S^3_{f\bar f}&=&{\textstyle {1\over 2}\int^{{\sqrt \tau}}_{\tau}{dx_2\over x_2}
\int_{x_2}^1dt_2\biggl({t_2^2+x_2^2\over t_2^2(t_2-x_2)_+}\ln {2x_2(x_2-\tau)
\over \tau(t_2+x_2)}+{3\over 2}{1\over (t_2-x_2)_+}-{2\over t_2}-
{3x_2\over t_2^2}\biggr)P_{f\bar f}(\tau/x_2,t_2,Q)},\nonumber \\
S^4_{f\bar f}&=&{\textstyle \int_{{\sqrt \tau}}^1{dx_1\over x_1}\int_{x_1}^1
dt_1\int_{\tau/x_1}^1dt_2{G^A(t_1,t_2)\over [(t_1-x_1)(t_2-\tau/x_1)]_+}
P_{f\bar f}(t_1,t_2,Q)},\nonumber \\
S^5_{f\bar f}&=&{\textstyle \int_{{\sqrt \tau}}^1{dx_1\over x_1}\int_{x_1}^1
dt_1\int_{\tau/x_1}^1dt_2{-2\tau (\tau+t_1t_2)\over t_1t_2(t_1\tau/x_1+
t_2x_1)^2}P_{f\bar f}(t_1,t_2,Q)}.
\label{fxss}
\end{eqnarray}
Eqs.~(\ref{fx1lpQ}) and (\ref{fxss}) were obtained using eqs. (2.6) and (C.1)
of \cite{ref:eightla}.  Note that the integration limits for $x_1, x_2$ follow
from the requirement that $y>0$ or equivalently, $x_F=x_1-x_2>0$.

In a similar way, starting from their eqs.~(2.17) and (C.1), the following
formula is obtained for the quark-gluon channel:
\begin{equation}
{d\sigma^{(1)}_{qg}\over dQ^2}\biggl|_{y>0}={2\pi \over 9}
{\alpha_e^2\alpha\over Q^2s}\sum_fe_f^2\sum_{j=1}^6 V^j_{gf},
\label{fx1lpQqg}
\end{equation}
with
\begin{eqnarray}
V^1_{gf}&=&{\textstyle \int_{\sqrt \tau}^1{dx_1\over x_1}\int_{x_1}^1dt_1
\biggl({x_1^2+(t_1-x_1)^2\over 2t_1^3}\ln{2x_1(x_1-\tau)\over \tau(t_1+x_1)}+{1
\over 2t_1}-{3x_1(t_1-x_1)\over t_1^3}
\biggr)P^C_{gf}(t_1,\tau/x_1,Q)},\nonumber \\
V^2_{gf}&=&{\textstyle \int^{\sqrt \tau}_\tau{dx_2\over x_2}\int_{\tau/x_2}^1
dt_1\int_{x_2}^1dt_2{G^C(t_1,t_2)\over (t_2-x_2)_+}P^C_{gf}(t_1,t_2,Q)},
\nonumber \\
V^3_{gf}&=&{\textstyle \int_{\sqrt \tau}^1{dx_1\over x_1}\int_{x_1}^1dt_1
\int_{\tau/x_1}^1dt_2H^C(t_1,t_2)P^C_{gf}(t_1,t_2,Q)},\nonumber \\
V^4_{gf}&=&{\textstyle \int^{\sqrt \tau}_\tau{dx_2\over x_2}\int_{x_2}^1dt_2
\biggl({x_2^2+(t_2-x_2)^2\over 2t_2^3}\ln{2x_2(x_2-\tau)\over \tau(t_2+x_2)}
+{1\over 2t_2}-{3x_2(t_2-x_2)\over t_2^3}\biggr)P^{C'}_{gf}(t_2,\tau/x_2,Q)},
\nonumber \\
V^5_{gf}&=&{\textstyle \int_{\sqrt \tau}^1{dx_1\over x_1}\int_{\tau/x_1}^1dt_2
\int_{x_1}^1dt_1{G^C(t_2,t_1)\over (t_1-x_1)_+}P^{C'}_{gf}(t_2,t_1,Q)},
\nonumber \\
V^6_{gf}&=&{\textstyle \int_{\sqrt \tau}^1{dx_1\over x_1}\int_{\tau/x_1}^1dt_2
\int_{x_1}^1dt_1H^C(t_2,t_1)P^{C'}_{gf}(t_2,t_1,Q)}.
\label{fxvs}
\end{eqnarray}

All of the various parton distributions used were taken from
the CERN PDFLIB package.

\newpage

\appendix{\bf FIGURE CAPTIONS}

\begin{description}
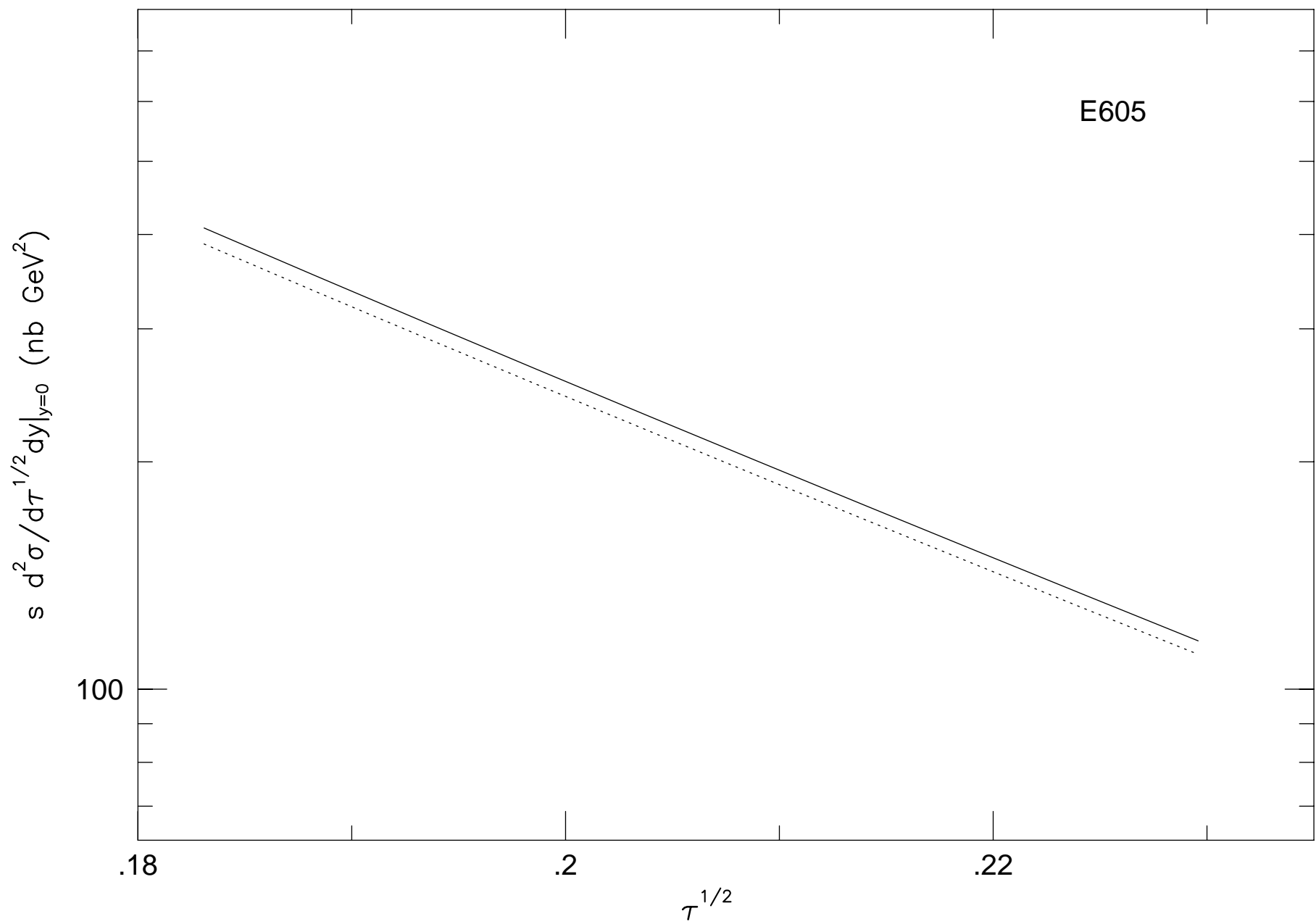
\item{Figure 1.} Exact and approximate $\displaystyle
{s{d^2\sigma \over d{\sqrt \tau}dy}
\biggl|_{y=0}} \ {\rm to} \ {\cal O}(\alpha)$.\\
Solid=exact calculation using results of Kubar et al in \cite{ref:eightla};\\
Dotted=calculated using Laenen-Sterman approximation.
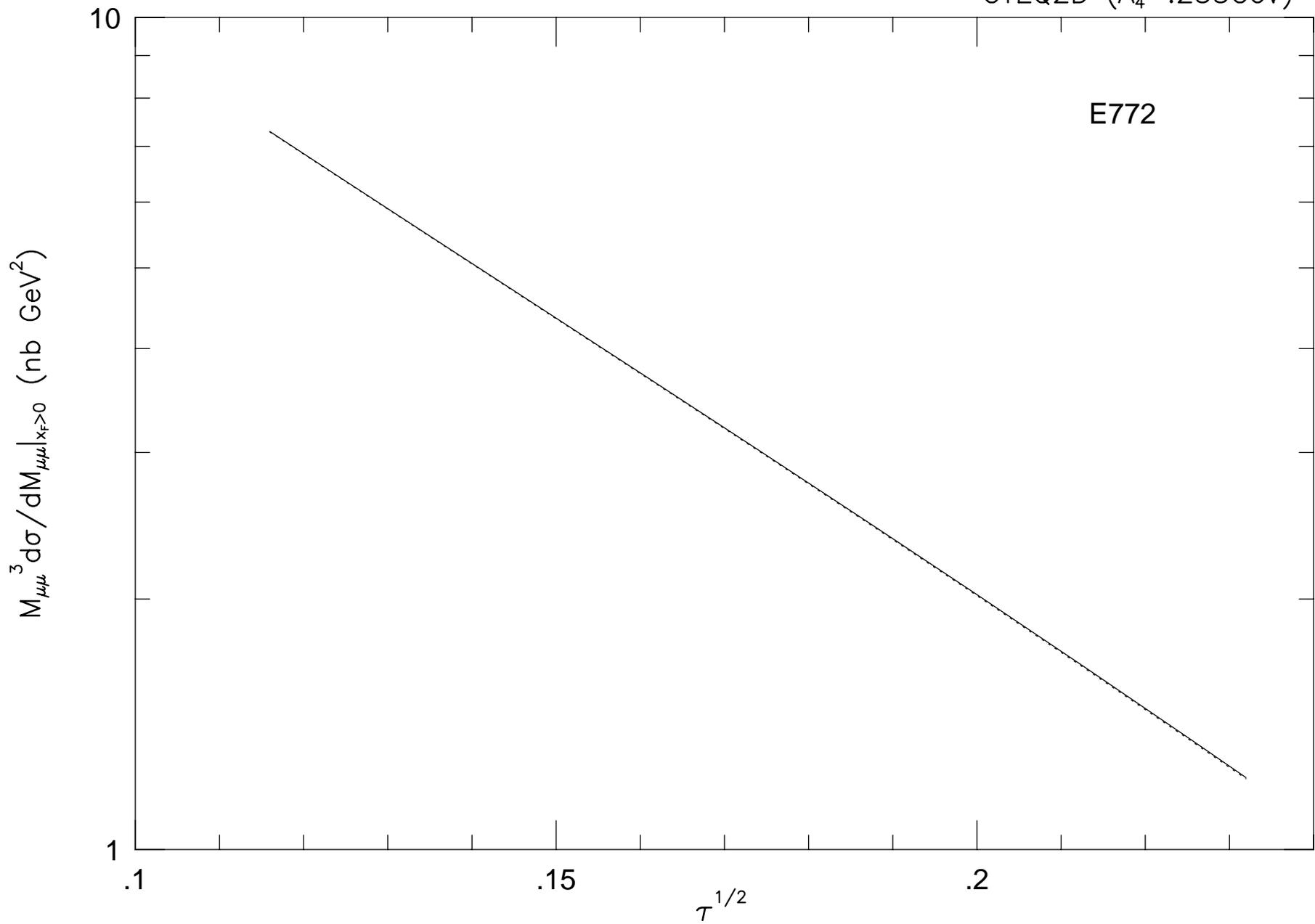
\item{Figure 2.} Exact and approximate $\displaystyle{M^3_{\mu\mu}{d\sigma
\over dM_{\mu\mu}}\biggl|_{x_F>0}} \
 {\rm to} \ {\cal O}(\alpha)$.\\
Solid=exact calculation using results of Kubar et al in \cite{ref:eightla};\\
Dotted=calculated using Laenen-Sterman approximation.
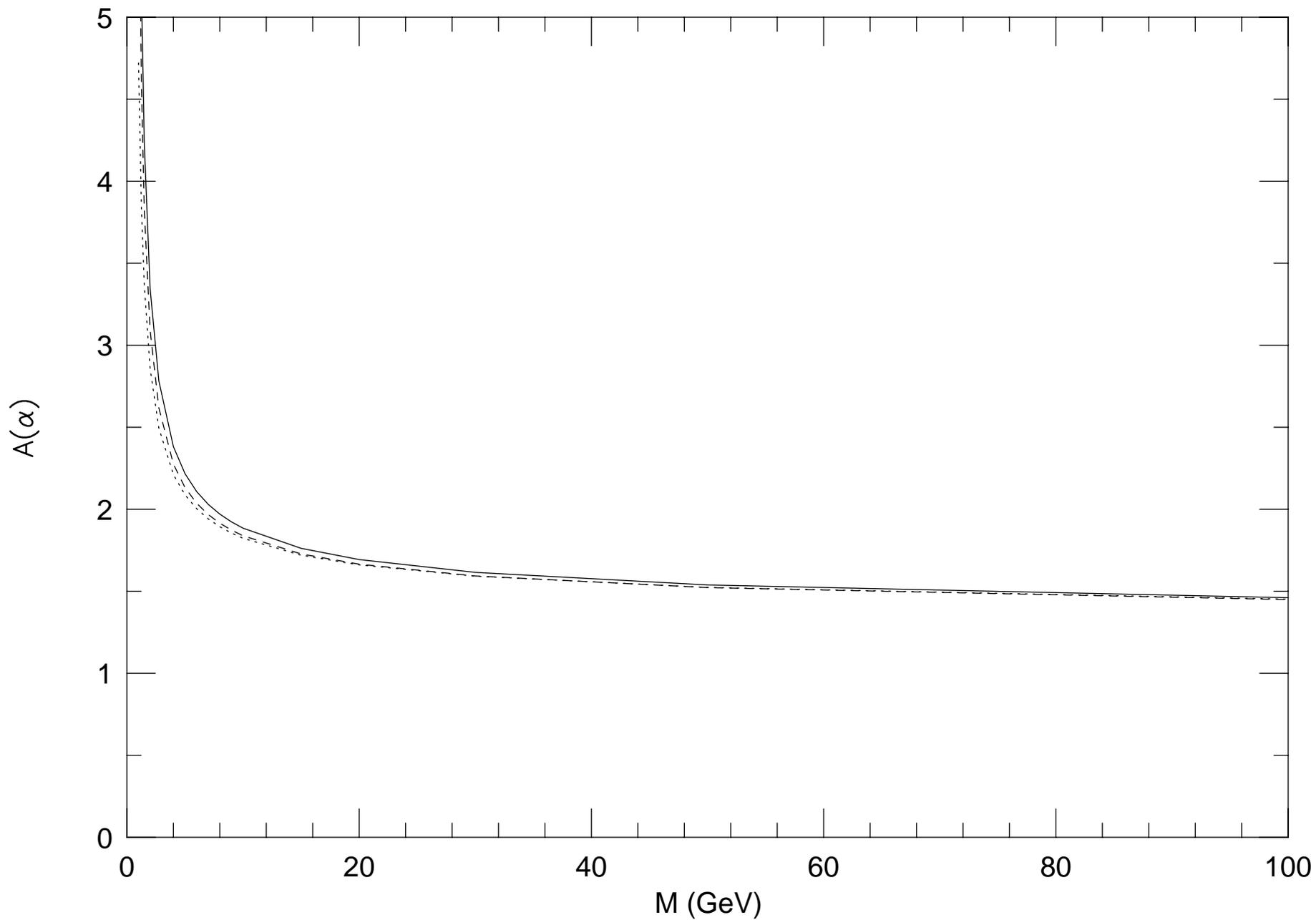
\item{Figure 3.} Different forms of $A(\alpha)$.\\
Solid=$A(\alpha)$; Dashed=$A_2(\alpha)$; Dotted=$A_1(\alpha)$.
\item{Figure 4.} NA3(1980): $\displaystyle {M^3_{\mu\mu} {d\sigma \over dM_{
\mu\mu}}\biggl|_{x_F>0}}$.
\begin{description}
\item{(a)} CTEQ2D.  Solid=resummed; Long Dashed=2-loop; Short Dashed=1-loop.
\item{(b)} Same as (a) but for MRS D-.
\end{description}
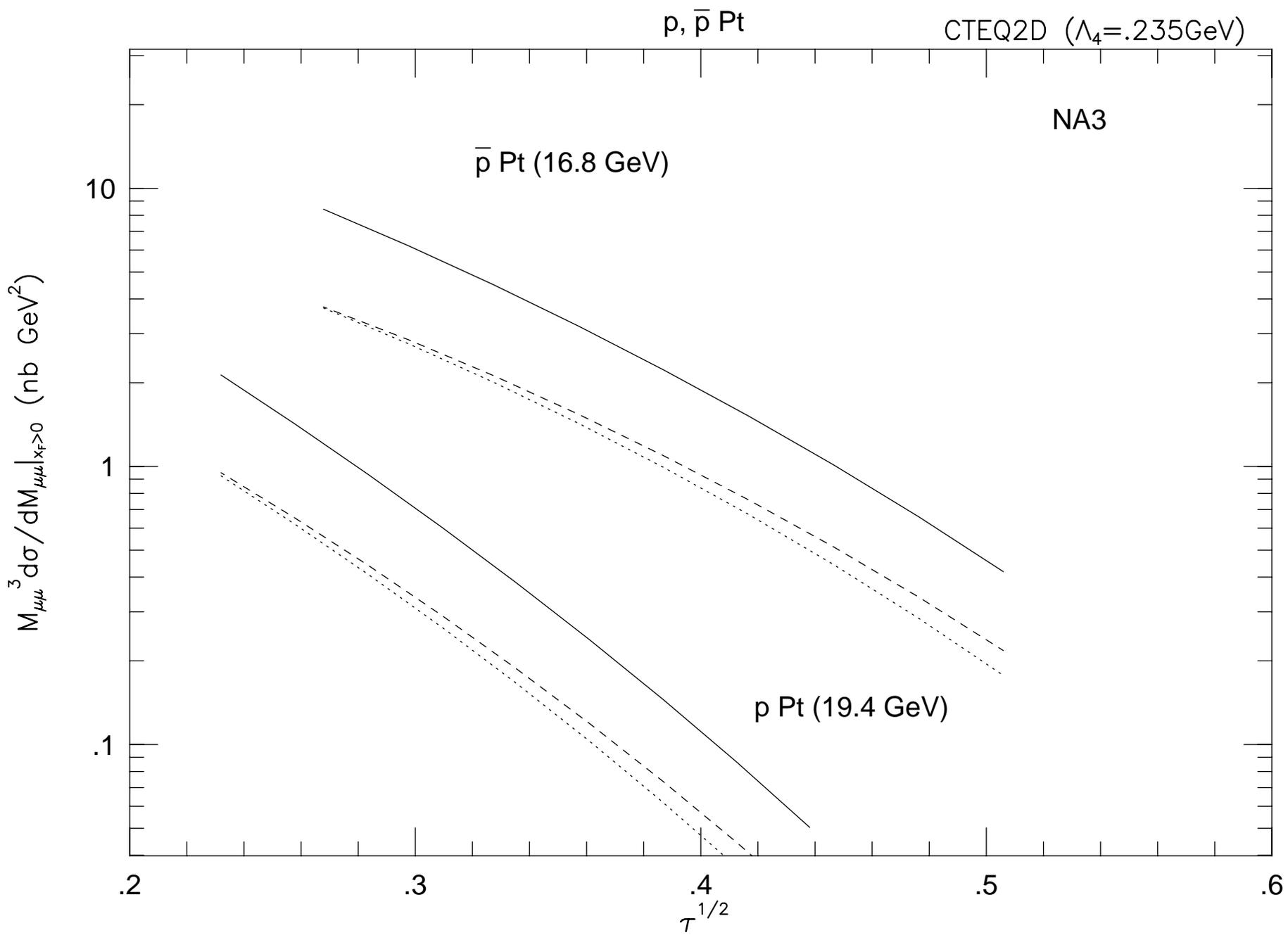
\item{Figure 5.} Contributions to $\displaystyle {M^3_{\mu\mu} {d\sigma
\over dM_{\mu\mu}}\biggl|_{x_F>0}}$.\\
Solid=total resummed cross section; Dashed=resummed but with $A(\alpha)$
replaced by 1;\\ Dotted=Born cross section.
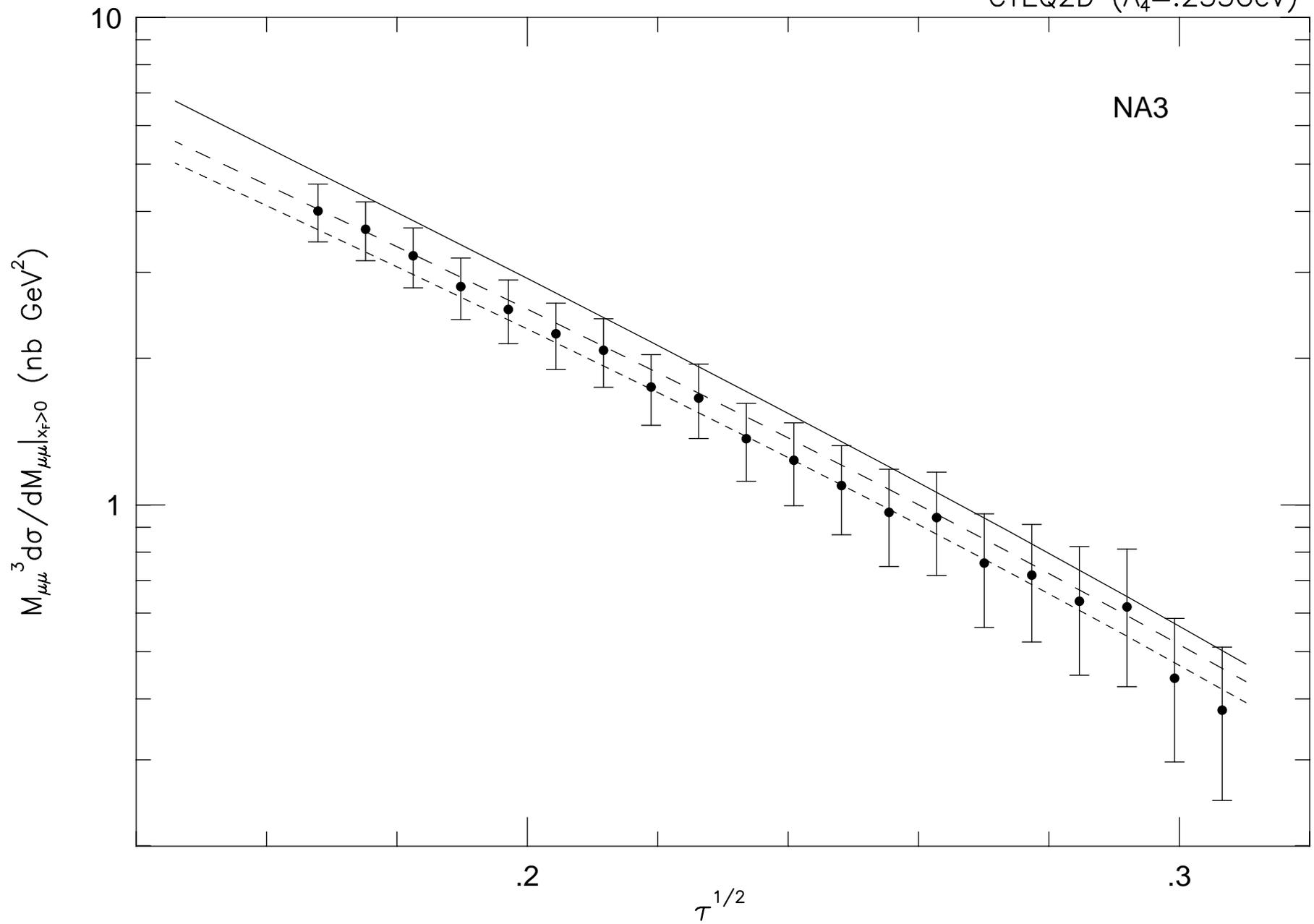
\item{Figure 6.} NA3(1985): $M^3_{\mu\mu}{\displaystyle {d\sigma\over
dM_{\mu\mu}}}\bigg|_{x_F>0}$ using CTEQ2D partons.\\
Solid=resummed; Long Dashed=2-loop; Short Dashed=1-loop.
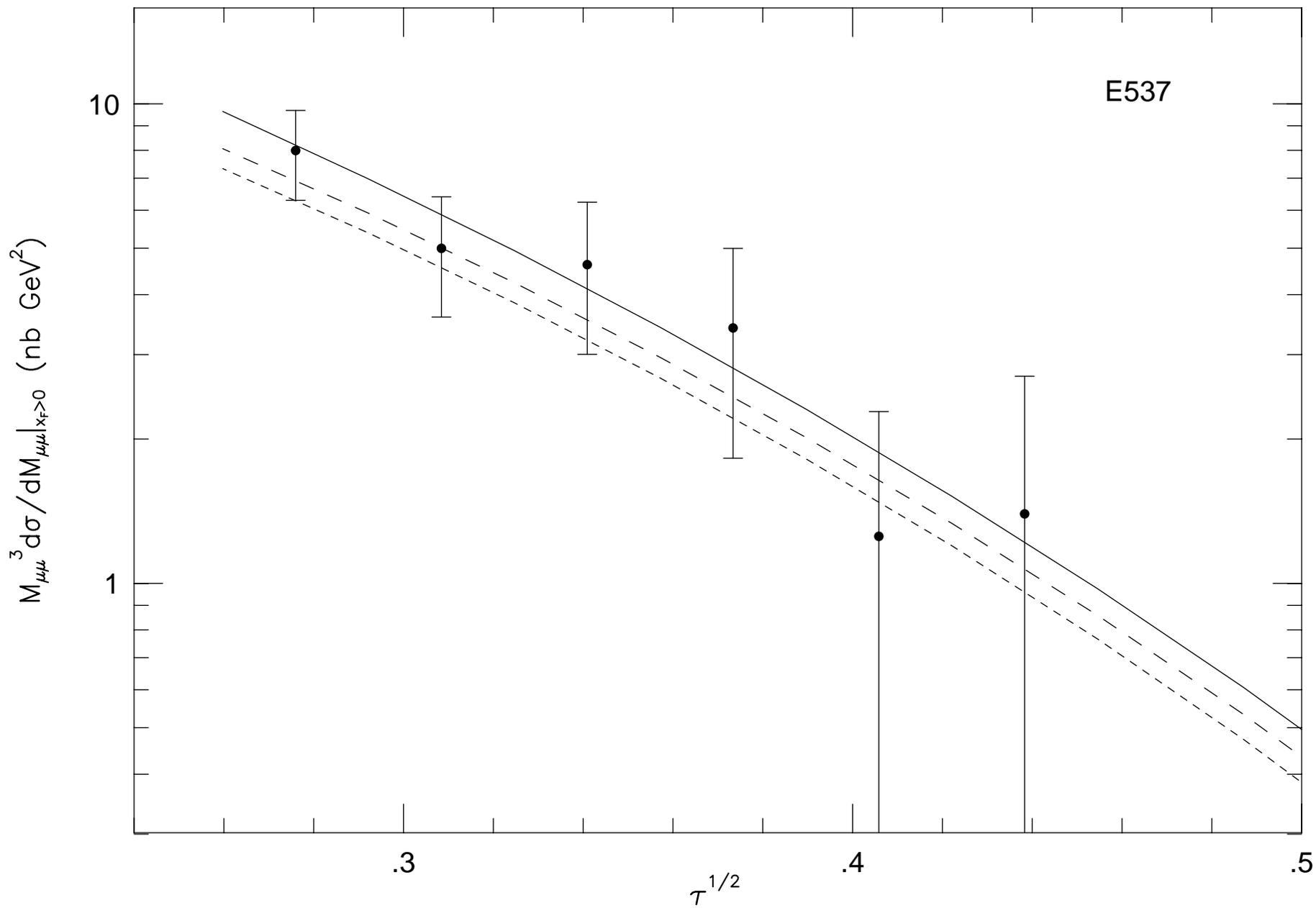
\item{Figure 7.} E537: $\displaystyle {M^3_{\mu\mu} {d\sigma \over dM_{
\mu\mu}}\biggl|_{x_F>0}}$ using CTEQ2D partons.\\
Solid=resummed; Long Dashed=2-loop; Short Dashed=1-loop.
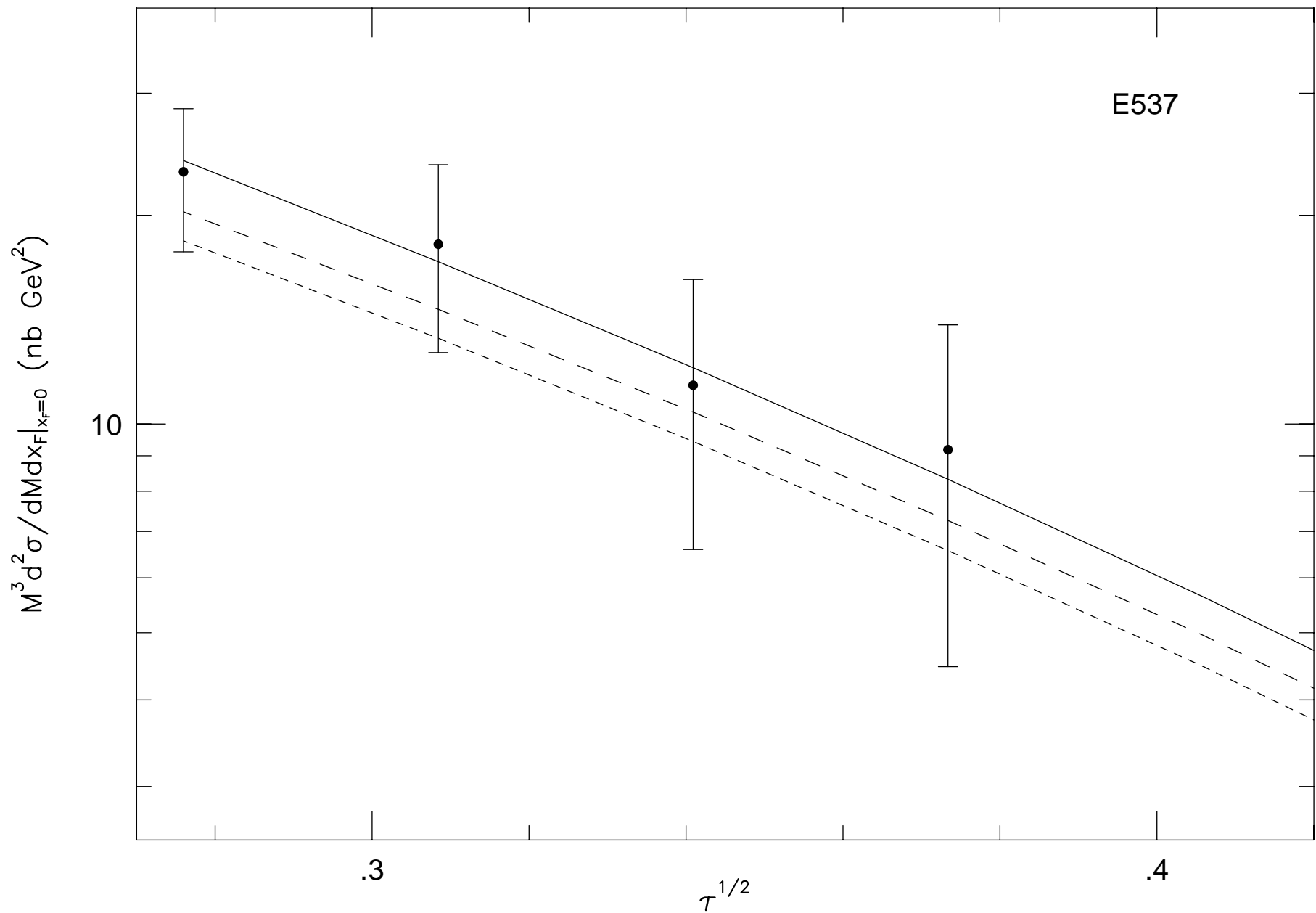
\item{Figure 8.} E537: $\displaystyle {M^3{d^2\sigma \over dMdx_F}
\biggl|_{x_F=0}}$ using CTEQ2D partons.\\
Solid=resummed; Long Dashed=2-loop; Short Dashed=1-loop.
\item{Figure 9.} E605: $\displaystyle {s {d^2\sigma \over d{\sqrt \tau}dy}
\biggl|_{y=0}}$.\\
Solid=resummed; Long Dashed=2-loop; Short Dashed=1-loop.
\begin{description}
\item{(a)} CTEQ2D.
\item{(b)} MRSD-.
\end{description}
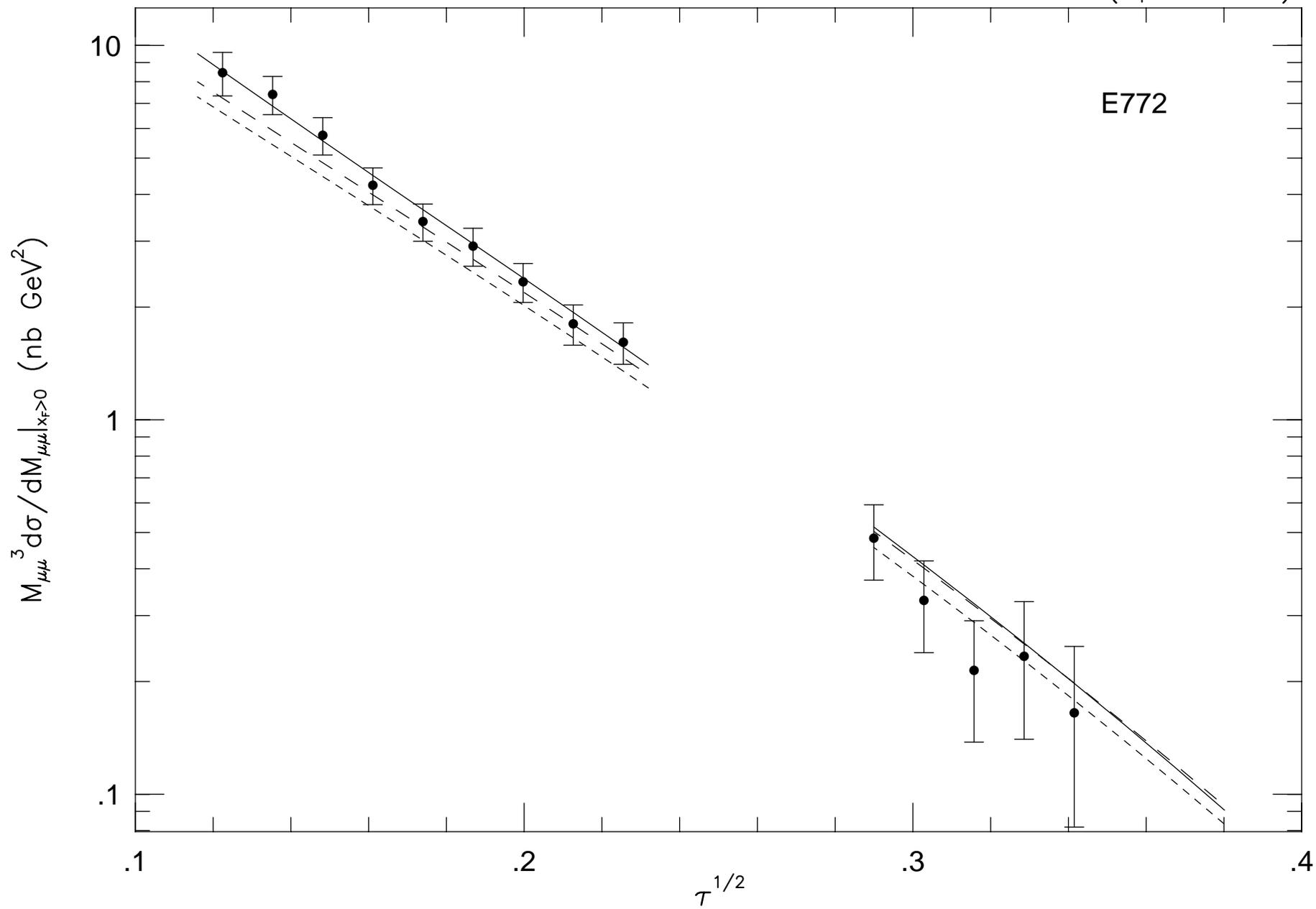
\item{Figure 10.} E772: $\displaystyle {M^3_{\mu\mu}{d\sigma\over dM_{\mu\mu}}}
\biggl|_{x_F>0}$ using CTEQ2D partons.\\
Solid=resummed; Long Dashed=2-loop; Short Dashed=1-loop.
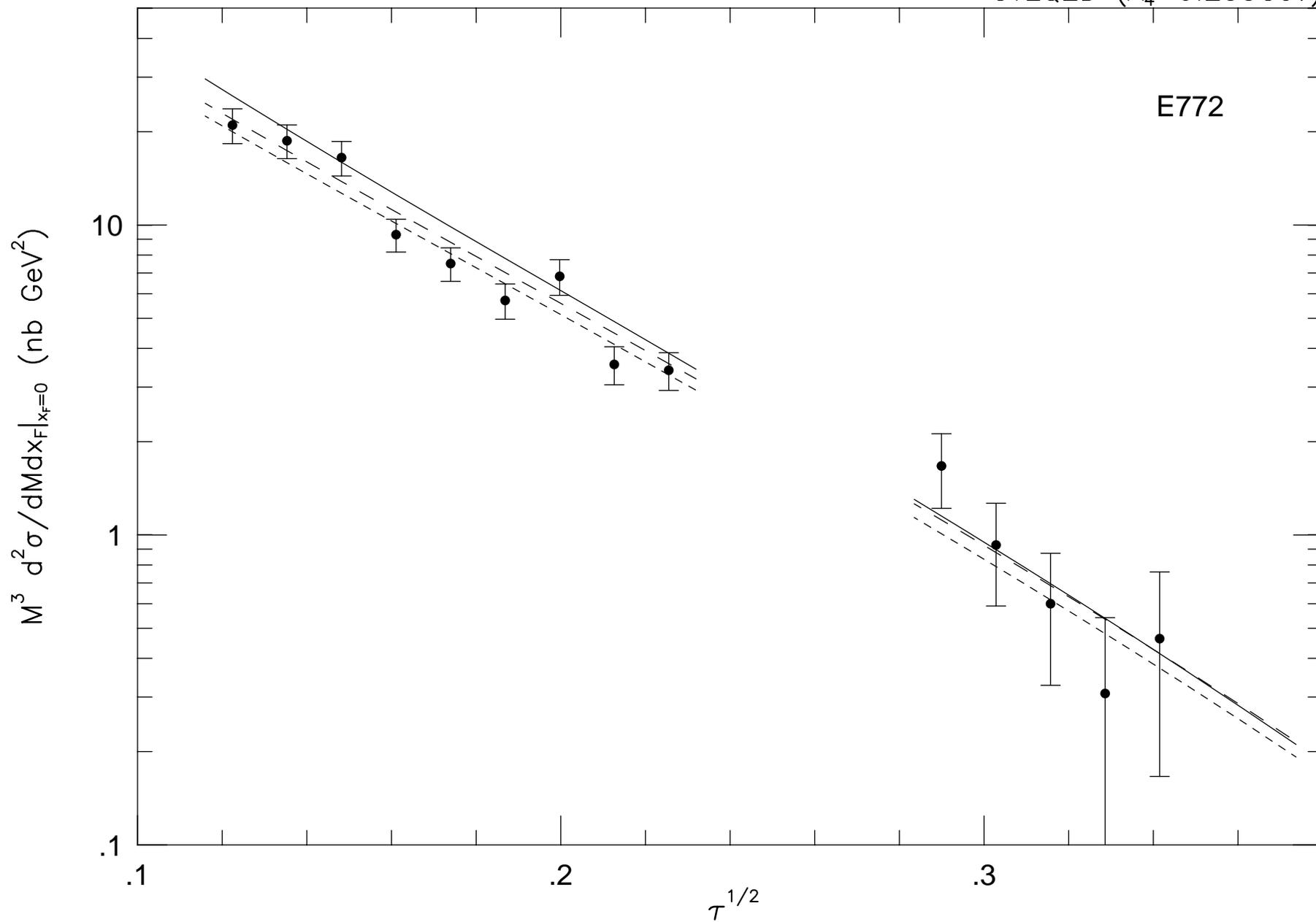
\item{Figure 11.} E772: $\displaystyle {M^3{d^2\sigma\over
dMdx_F}}\biggl|_{x_F=0}$ using CTEQ2D partons.\\
Solid=resummed; Long Dashed=2-loop; Short Dashed=1-loop.
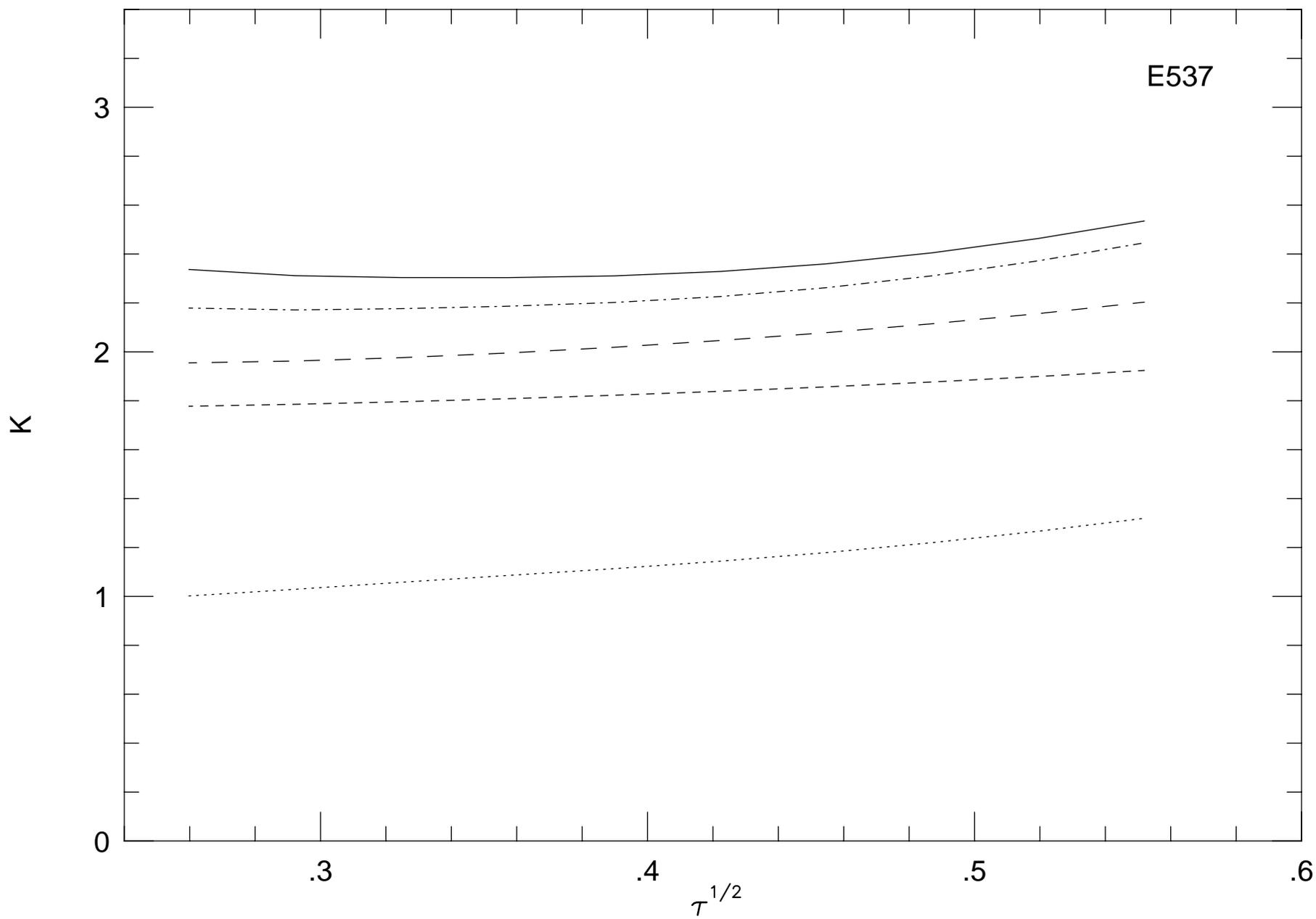
\item{Figure 12.} $K$ factors for E537 $\bar p$-W using CTEQ2D.\\
Solid=resummed; Long Dashed=2-loop; Short Dashed=1-loop;\\
Dot-Dashed=resummed with $A(\alpha)$ replaced by $A_1(\alpha)$;\\
Dotted=resummed with $A(\alpha)$ replaced by $1$.
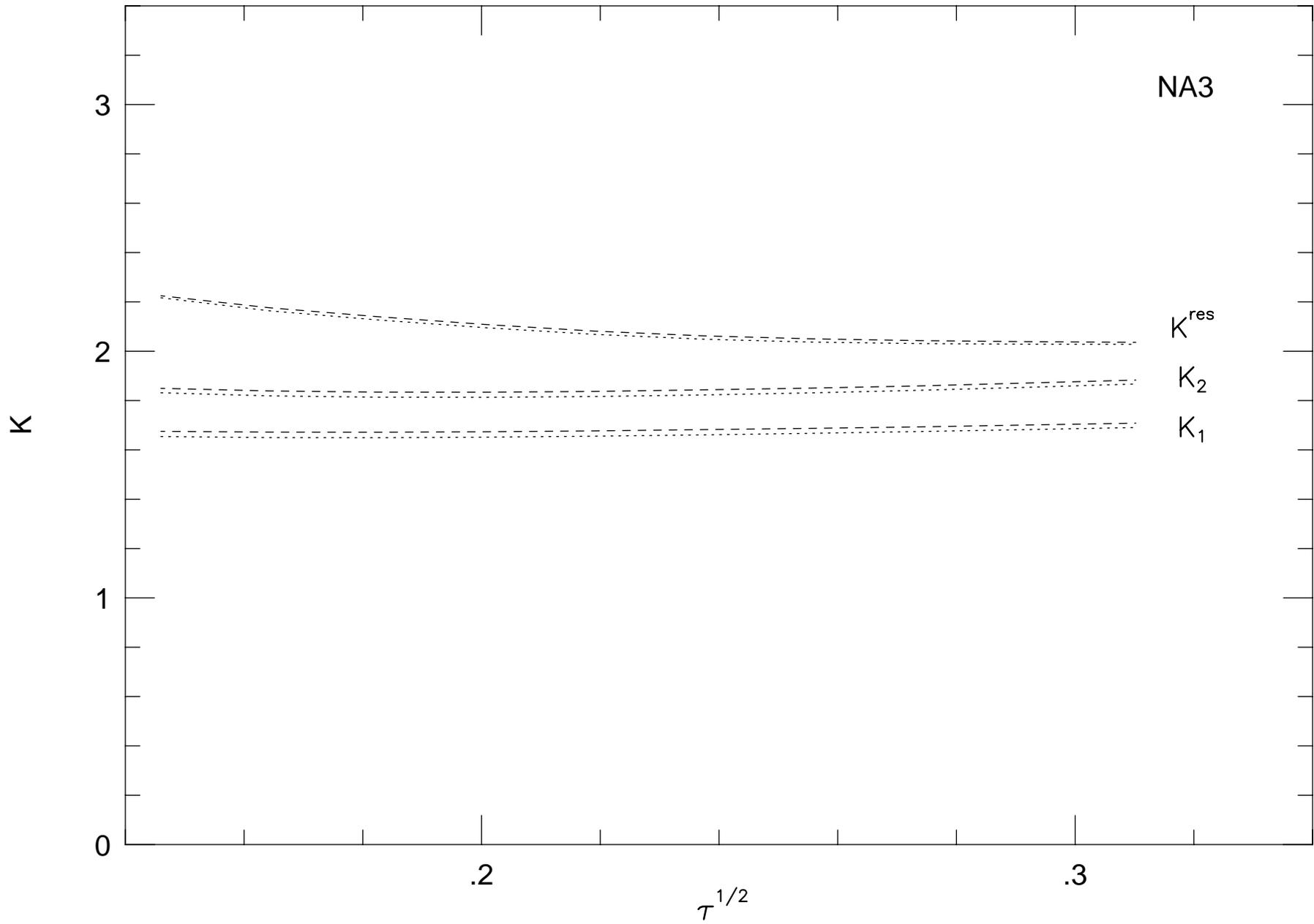
\item{Figure 13.} NA3(1985) theoretical $K$-factors for
different parton sets.\\
Dashed (descending order): Same as above using MRSD-;\\
Dotted (descending order): Same as above using CTEQ2D.
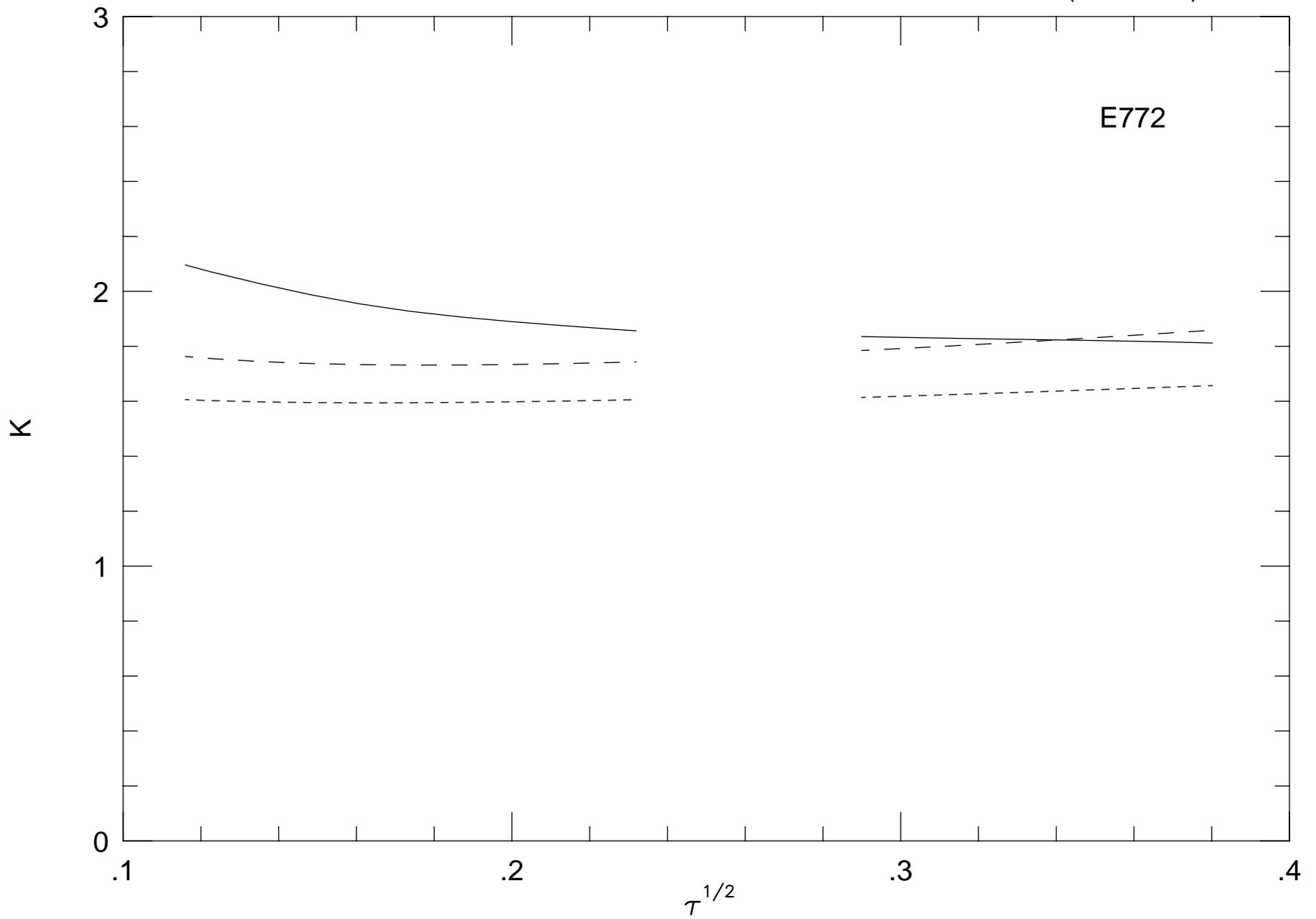
\item{Figure 14.} $K$-factors for E772 $p$-$^2$H using CTEQ2D.\\
Solid=resummed; Long Dashed=2-loop; Short Dashed=1-loop.
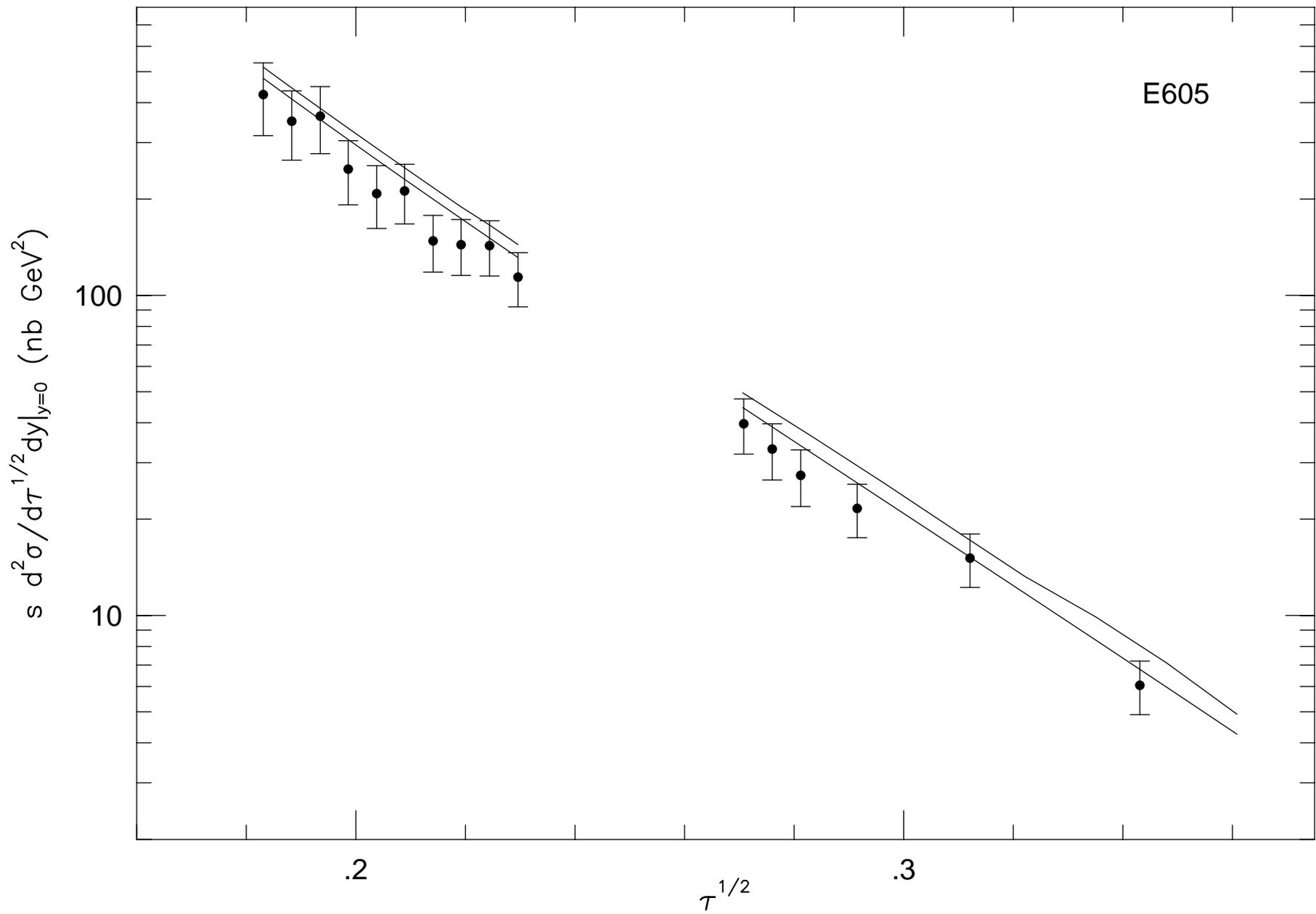
\item{Figure 15.} Resummed curves for E605 using global fits.\\
Upper solid curve=MRSD-; Lower solid curve=CTEQ2D.
\end{description}


\begin{thebibliography}{99}
\bibitem{ref:populace}
G. Altarelli, R.K. Ellis and G. Martinelli, Nucl. Phys. B157 (1979) 461;\\
 B. Humpert and W.L. van Neerven, Phys. Lett. B84 (1979) 327;\\
J. Kubar-Andre and F.E. Paige, Phys. Rev. D19 (1979) 221; \\ K. Harada,
T. Kaneko and N. Sakai, Nucl. Phys. B155 (1979) 169; B165 (1980) 545 (E); \\
T. Matsuura and W. L. van Neerven, Z. Phys. C38 (1988) 623; \\
T. Matsuura, S.C. van der Marck and W.L. van Neerven, Phys. Lett. B211 (1988)
171.
\bibitem{ref:parti}
L. Alvero and H. Contopanagos, Nucl. Phys. B436 (1995) 184.
\bibitem{ref:paptwo}
H. Contopanagos and G. Sterman, Nucl. Phys. B419 (1994) 77.
\bibitem{ref:six}
D. Appell, P. Mackenzie and G. Sterman, Nucl. Phys. B309 (1988) 259.
\bibitem{ref:vanNeervendis}
T. Matsuura, S.C. van der Marck and W.L. van Neerven, Nucl. Phys. B319 (1989)
570.
\bibitem{ref:three}
L. Magnea and G. Sterman, Phys. Rev. D42 (1990) 4222.
\bibitem{ref:one}
R. Hamberg,
W.L. van Neerven and T. Matsuura, Nucl. Phys. B359 (1991) 343.
\bibitem{ref:two}
CTEQ collaboration
 (G. Sterman, editor) {\it Handbook of perturbative QCD}, version 1.0
(April 1993).
\bibitem{ref:papzero}
H. Contopanagos and G. Sterman, Nucl. Phys. B400 (1993) 211.
\bibitem{ref:vanNeervendisp}
W.L. van Neerven and E.B. Zijlstra, Nucl. Phys. B382 (1992) 11.
\bibitem{ref:vanNeervendispp}
P.J. Rijken and W.L. van Neerven, Phys. Rev. D51 (1995) 44.
\bibitem{ref:laester}
E. Laenen and G. Sterman, {\it The Fermilab Meeting}, DPF'92, Vol. 2,
987 (C. Albright et al., editors).
\bibitem{ref:eightla}
J. Kubar, M. Le Bellac, J.L. Meunier and G. Plaut, Nucl. Phys. B175 (1980) 251.
\bibitem{ref:cteq}
CTEQ Collaboration, unpublished; H.L. Lai et al., MSU-HEP-41024/CTEQ-404
(October 1994).
\bibitem{ref:mrsd}
A.D. Martin, W.J. Stirling and R.G. Roberts, Phys. Rev. D47 (1993) 867;
Phys. Lett. 306B (1993) 145.
\bibitem{ref:nathree}
J. Badier et al., Phys. Lett. B96 (1980) 422.
\bibitem{ref:nathreep}
J. Badier et al., Z. Phys. C26 (1985) 489.
\bibitem{ref:efive}
E. Anassontzis et al., Phys. Rev. D38 (1988) 1377.
\bibitem{ref:esix}
G. Moreno et al., Phys. Rev. D43 (1991) 2815.
\bibitem{ref:eseven}
P.L. McGaughey et al. (E772 Collaboration), Phys. Rev. D50 (1994) 3038.
\bibitem{ref:LA}
L. Alvero, ITP-SB-94-67 (December 1994), hep-ph 9412335.
\bibitem{ref:dflm}
M. Diemoz, F. Ferroni, E. Longo and G. Martinelli, Z. Phys. C39 (1988) 21.
\bibitem{ref:ehlq}
E. Eichten, I. Hinchliffe, K. Lane and C. Quigg, Rev. Mod. Phys. 56 (1984) 579.
\bibitem{ref:gs87}
G. Sterman, Nucl. Phys. B281 (1987) 310;\\
S. Catani and L. Trentadue, Nucl. Phys. B327 (1989) 323; B353 (1991) 183.
\bibitem{ref:top}
E. Berger and H. Contopanagos, ANL-HEP-PR-95-31 (July 1995).
\end{thebibliography}
\end{document}